\documentclass[fleqn,usenatbib]{mnras}

\usepackage{newtxtext,newtxmath}

\usepackage[T1]{fontenc}
\usepackage{ae,aecompl}

\usepackage{graphicx}	
\usepackage{amsmath}	

\usepackage{lscape}
\usepackage{rotating}
\usepackage{tablefootnote}
\usepackage{xcolor}

\title[Multi-wavelength properties of the NEPSC2 sources]{Multi-wavelength properties of 850-$\mu$m selected sources from the North Ecliptic Pole SCUBA-2 survey}

\author[H. Shim et al.]
{Hyunjin Shim,$^{1}$\thanks{E-mail: hjshim@knu.ac.kr}
Dongseob Lee,$^{1}$
Yeonsik Kim,$^{2}$
Douglas Scott,$^{3}$
Stephen Serjeant,$^{4}$
\newauthor
Yiping Ao,$^{5,6}$
Laia Barrufet,$^{7,8}$
Scott C. Chapman,$^{3,9}$
David L. Clements,$^{10}$
\newauthor
Christopher J. Conselice,$^{11}$
Tomotsugu Goto,$^{12}$
Thomas R. Greve,$^{13,14,15}$
\newauthor
Ho Seong Hwang,$^{16,17}$ 
Myungshin Im,$^{17}$ 
Woong-Seob Jeong,$^{18}$
Helen K. Kim,$^{19}$
\newauthor
Minjin Kim,$^{2}$
Seong Jin Kim,$^{12}$ 
Albert K. H. Kong,$^{12}$
Maciej P. Koprowski,$^{20}$
\newauthor
Matthew A. Malkan,$^{19}$
Micha{\l}~J.~Micha{\l}owski,$^{21}$
Chris Pearson,$^{4,8,15}$
Hyunjong Seo,$^{18}$ 
\newauthor
Toshinobu Takagi,$^{22}$ 
Yoshiki Toba,$^{23,24,25}$
Glenn J. White,$^{4,8}$
and Jong-Hak Woo$^{16}$
\\
\\
$^{1}$Department of Earth Science Education, Kyungpook National University, Daegu 41566, Republic of Korea\\
$^{2}$Department of Astronomy and Atmospheric Sciences, Kyungpook National University, Daegu 41566, Republic of Korea\\
$^{3}$Department of Physics and Astronomy, University of British Columbia, Canada\\
$^{4}$The Open University, Milton Keynes, MK7 6AA, UK\\
$^{5}$Purple Mountain Observatory and Key Laboratory for Radio Astronomy, Chinese Academy of Sciences, Nanjing, China\\
$^{6}$School of Astronomy and Space Science, University of Science and Technology of China, Hefei, Anhui, China\\
$^{7}$Geneva Observatory, University of Geneva, Versoix, Switzerland  \\
$^{8}$RAL Space, STFC Rutherford Appleton Laboratory, Didcot, Oxfordshire OX11 0QX, UK \\
$^{9}$Department of Physics and Atmospheric Science, Dalhouise University, Halifax, NS B3H 4R2, Canada \\
$^{10}$Imperial College London, Blackett Lab., Prince Consort Road, London, SW7 2AZ, UK\\
$^{11}$Department of Physics and Astronomy, University of Manchester, UK \\ 
$^{12}$Institute of Astronomy, National Tsing Hua University, 101, Section 2. Kuang-Fu Road, Hsinchu, 30013, Taiwan (R.O.C.) \\
$^{13}$Cosmic Dawn Center \\
$^{14}$National Space Institute, DTU Space, Technical University of Denmark, Elektrovej 327, DK-2800 Kgs.~Lyngby, Denmark\\
$^{15}$Department of Physics and Astronomy, University College London, Gower Street, London WC1E 6BT, UK \\
$^{16}$Astronomy Program, Department of Physics and Astronomy, Seoul National University, 1 Gwanak-ro, Gwanak-gu, Seoul 08826, Republic of Korea\\
$^{17}$SNU Astronomy Research Center, Astronomy Program, Department of Physics and Astronomy, Seoul National University, Republic of Korea\\
$^{18}$Korea Astronomy and Space Science Institute, Daejeon 34055, Republic of Korea\\
$^{19}$Department of Physics and Astronomy, UCLA, Los Angeles, CA 90095-1547, USA\\
$^{20}$Institute of Astronomy, Faculty of Physics, Astronomy and Informatics, Nicolaus Copernicus University, Grudziadzka 5, 87-100 Torun, Poland\\
$^{21}$Astronomical Observatory Institute, Faculty of Physics, Adam
Mickiewicz University, ul.~S{\l}oneczna 36, 60-286 Pozna{\'n}, Poland \\
$^{22}$Japan Science Forum, 3-2-1, Kandasurugadai, Chiyoda-ku, Tokyo 101-0062, Japan\\
$^{23}$Department of Astronomy, Kyoto University, Kitashirakawa-Oiwake-cho, Sakyo-ku, Kyoto 606-8502, Japan\\
$^{24}$Academia Sinica Institute of Astronomy and Astrophysics, 11F of Astronomy-Mathematics Building, AS/NTU, No.1, Section 4, Roosevelt Road, Taipei 10617, Taiwan\\ 
$^{25}$Research Center for Space and Cosmic Evolution, Ehime University, 2-5 Bunkyo-cho, Matsuyama, Ehime 790-8577, Japan\\
}

\date{Accepted XXX. Received YYY; in original form ZZZ}

\pubyear{2022}

\begin{document}
\label{firstpage}
\pagerange{\pageref{firstpage}--\pageref{lastpage}}
\maketitle
\clearpage

\begin{abstract}
	We present the multi-wavelength counterparts 
	of 850-$\mu$m selected submillimetre sources 
	over a 2-deg$^2$ field centred on the North Ecliptic Pole.
	In order to overcome the large beam size (15\,arcsec) of the 
	850-$\mu$m images, 
	deep optical to near-infrared (NIR) photometric data 
	and arcsecond-resolution 20-cm images are used
	to identify counterparts of submillimetre sources.
	Among 647 sources, we identify
	514 reliable counterparts for 449 sources (69 per cent in number), 
	based either on probabilities of chance associations
	calculated from positional offsets 
	or offsets combined with the optical-to-NIR colours.  
	In the radio imaging, the fraction of 850-$\mu$m sources 
	having multiple counterparts is 7 per cent.
	The photometric redshift, infrared luminosity, 
	stellar mass, star-formation rate (SFR), 
	and the AGN contribution to the total infrared luminosity 
	of the identified counterparts 
	are investigated through spectral energy distribution fitting.
	The SMGs are infrared-luminous galaxies 
	at an average $\langle z\rangle=2.5$
	with $\mathrm{log}_{10} (L_\mathrm{IR}/\mathrm{L}_\odot)=11.5$--13.5,
	with a mean stellar mass of 
	$\mathrm{log}_{10} (M_\mathrm{star}/\mathrm{M}_\odot)=10.90$ 
	and SFR of $\mathrm{log}_{10} (\mathrm{SFR/M_\odot\,yr^{-1}})=2.34$. 
	The SMGs show twice as large SFR as galaxies on the star-forming main sequence, 
	and about 40 per cent of the SMGs are classified as 
	objects with bursty star formation. 
	At $z\ge4$, the contribution of AGN luminosity to total luminosity 
	for most SMGs is larger than 30 per cent. 
	The FIR-to-radio correlation coefficient of SMGs is
	consistent with that of main-sequence galaxies at $z\simeq2$.
\end{abstract}




\section{Introduction}

The population of Submillimetre galaxies 
\citep[hereafter SMGs;][]{Smail1998, Blain2002}
has been unveiled and widely studied 
during the past two decades, 
supporting the idea that a significant fraction of
star formation is located in heavily dust-obscured galaxies 
during the peak epoch of cosmic star formation. 
Early surveys of SMGs were carried out
using the low opacity atmospheric windows 
at 450\,$\mu$m and 850\,$\mu$m
\citep[e.g.,][]{Smail1997, Barger1998}. 
The subsequent space-based observatory \textit{Herschel}
observations greatly increased the number of galaxies 
selected in 250, 350, and 500\,$\mu$m 
\citep[e.g.,][]{Bourne2016, Ward2022}. 
Follow-up studies have shown that 
SMGs are gas-rich \citep{Tacconi2006, Riechers2010}
and massive \citep{Michalowski2012, Michalowski2014, Dudzeviciute2020}
galaxies with large ($100$--$1000$\,M$_{\odot}$\,yr$^{-1}$)
star formation rates \citep[SFRs;][]{Barger2014, Toba2020a}
and heavily attenuated by dust  
\citep{Geach2007, Michalowski2010}.
SMGs are frequently associated with an active galactic nucleus
\citep[AGN;][]{Wang2013, Toba2018, Ueda2018},
although the AGN component rarely dominates
the submillimetre emission \citep[e.g.,][]{Laird2010}.
Since the contribution of infrared-luminous galaxies
to the global SFR density
increases continuously up to
$z=2$ and 3 \citep[e.g.,][for a review]{Magnelli2013, Lim2020, Casey2014},
understanding the physical nature of SMGs  
is crucial for constructing a self-consistent galaxy evolution model. 

Statistical quantities, e.g., 
the number counts of 850-$\mu$m selected galaxies 
at a flux limit of $S_{850}\,{>}\,1\,\mathrm{mJy}$
have now reached overall agreement 
between a variety of degree-scale survey fields
\citep[][]{Geach2017, Simpson2020}.
Recent galaxy formation models
also provide relatively good fits
to the observed number counts
and/or FIR luminosity function of SMGs
\citep[e.g.,][]{Lagos2020, Lovell2021}
without requiring special treatment
such as a different form of the initial mass function.
This supports the idea that many SMGs represent
a natural evolutionary stage of typical galaxies
rather than being an exceptional galaxy population.
Nevertheless, there remains a debate about the variation of
physical properties for individual SMGs,
i.e., why some SMGs lie 
above the so-called star-forming main sequence of galaxies
\citep[e.g.,][]{daCunha2015} while others do not. 
As the number counts of SMGs can used to
constrain the parameters for the galaxy formation models
\citep[e.g.,][]{Cowley2015},
other statistics, such as
the distribution of SFR, stellar mass, dust-to-gas mass ratio,
redshift and AGN luminosity of SMGs,
also provide strong constraints on the models
\citep{Hayward2021}. However, 
despite the success in fitting number counts,
it is still difficult to reproduce all
the observed physical properties of SMGs
in large volume surveys with galaxy formation models,
due to the uncertainties
in the fueling and feedback mechanisms
that lead to the growth of SMGs.

Earlier ideas 
to explain active star formation in SMGs 
include merger-induced star formation 
\citep{Narayanan2009, Narayanan2010},
similar to the case of 
local ultra-luminous infrared galaxies 
\citep[ULIRGs;][]{Sanders1988}. 
Since gas-rich galaxy-galaxy major merger events 
may cause both 
enhanced star formation and elevated AGN activity
\citep{Hopkins2008}, 
diverse high-redshift galaxy populations 
may represent different phases of galaxy evolution. 
In the framework of gas-rich mergers for 
describing the formation of massive galaxies, 
SMGs are suspected to represent 
a starburst-dominated phase 
in the earlier stages of evolution, 
while the later quasar-dominated phase 
is observed as dust-obscured AGN 
\citep{Dey2008, Narayanan2010, HwangGeller2013, Toba2015, Toba2016}. 
10 to 20 per cent of SMGs are suspected to host AGN 
based on X-ray detection \citep{Johnson2013}. 
The fraction of AGN among SMGs \citep{Wang2013}
and the contribution of AGN 
to the bolometric luminosity of SMGs \citep{Pope2008, Seo2018}
provide indirect probes of 
the massive galaxy evolution scenario 
that includes an SMG phase. 
By probing black-hole-mass to stellar-mass ratios in SMGs, 
it is reported that SMGs host less massive black holes 
compared to non-SMGs with the same stellar mass 
at $z<2$
\citep{Alexander2008, Ueda2018}, 
supporting the suggestion that black hole growth comes 
after the growth of stellar mass. 
On the other hand, it is still an open question 
whether the star formation is enhanced or 
quenched through AGN activity in SMGs 
\citep[e.g.,][]{Ramasawmy2019},
which is related to the more general issue 
of the strength of AGN feedback in galaxy evolution. 

Another possibility is that star formation in SMGs 
is powered by less vigorous, continuous fueling
of gas through gas accretion or infall
\citep{Dave2010}.
Recent galaxy-formation models show that 
the submillimetre-bright phase can last
for long enough to allow a supply of infalling gas, 
without requiring a major merger 
\citep{Narayanan2015}. 
From these simulated galaxies, 
the merger fraction of SMGs is not significantly higher 
compared with non-SMGs, suggesting that mergers are 
not the main origin of the high SFRs in SMGs 
\citep{McAlpine2019}.
Direct studies of stellar and gas morphologies
would quantify the fraction of
merger-driven starbursts in the SMG population.
However, this requires deep, high-resolution
imaging follow up.
While such observations are time-consuming, 
an alternative way
to distinguish between mergers 
and continuous gas infall 
is to use the star-forming main sequence 
revealed in an SFR versus stellar mass diagram 
\citep[e.g.,][]{Speagle2014, Schreiber2015, Koprowski2016, Tomczak2016}. 
Although there is still debate about 
whether there exists a mass dependency 
in the specific SFR, 
the observed star-forming main sequence 
and its evolution as a function of redshift 
are well described by some recent galaxy-formation models 
\citep[e.g.,][]{Henriques2015, Nelson2015}. 
Therefore, the location of a galaxy 
relative to the star-forming main sequence
provides information about
the star-formation efficiency, 
which could be related to mergers
\citep[e.g.,][]{Barrufet2020}.
This idea is also supported by radio observations 
such as those described by \citet{Miettinen2017},
which shows that 
SMGs on the main sequence are disk-like galaxies 
with relatively large radio sizes, 
while SMGs above the main sequence appear to be irregular
in the rest-frame UV.
A complication is that the placement with respect to the main sequence 
seems to depend on the method used to derive stellar mass, 
as discussed by \citet{Michalowski2012, Michalowski2014}.

In order to probe 
the formation and evolution of SMGs, it is necessary to
identify SMGs in other wavebands. 
For instance, an optical/near-infrared (NIR) identification is crucial 
for estimating properties such as stellar mass 
and photometric redshift. 
Unfortunately, SMGs selected in wide-area surveys
suffer from the large beam sizes of single-dish imaging, 
which is of the order of 10\,arcsec. 
This value is far larger than the beam size of 
the optical and NIR data, 
which is of order 1\,arcsec. 
Therefore, since more than one optical/NIR selected object 
can fall within the beam size of submillimetre image, 
it is far from straightforward 
to determine which objects are responsible for the 
observed submillimetre emission. 
The use of submillimetre/millimetre interferometer array systems 
for follow up of discovered submillimetre sources 
\citep[e.g.,][]{Karim2013, Hill2018, Simpson2020, Dudzeviciute2020} 
is required to obtain a better spatial localization of 
submillimetre source counterparts. 
With the more precise coordinates, 
it is possible to probe the physical properties of SMGs 
including securing redshift estimation. 
If no higher spatial resolution images are available 
in the submillimetre/millimetre,
radio images can be utilized instead, 
because of the tight FIR-radio correlation 
for star-forming galaxies \citep[e.g.,][]{Condon1992}. 
Such radio identification has also been frequently used 
for identification of submillimetre sources 
\citep[e.g.,][]{Ivison2002, Chapman2001, Chapman2003, Smolcic2015, Liu2019, Lim2020}. 
Likelihood methods using positions, magnitudes and/or
colours (from the deep multi-wavelength data sets) 
have also been developed 
to identify counterparts of submillimetre sources 
to overcome the lack of high spatial resolution data 
at wavelengths longer than the submillimetre 
\citep[e.g.,][]{Chapin2009, Hwang2010, Casey2013, Michalowski2017, An2019}.
This suggests that the areas with rich multi-wavelength 
data available are ideal
for studying the physical properties of submillimetre-detected sources.

The James Clerk Maxwell Telescope (JCMT)
Submillimetre Common User Bolometer Array-2 
\citep[SCUBA-2;][]{Holland2013} 
Cosmology Legacy Survey (S2CLS) 
has been initiated as an 850-$\mu$m survey over 
several frequently visited fields \citep{Geach2017} 
with just the above motivation.
About 3000 850-$\mu$m sources are detected
in the 5\,deg$^2$ of S2CLS fields.
Detailed analyses of the physical properties of SMGs 
in S2CLS 
have been conducted on the EGS \citep{Zavala2018},
COSMOS \citep{Michalowski2017, An2019}, 
and UKIDSS/UDS 
\citep{Chen2016, Michalowski2017, Dudzeviciute2020}
fields, 
which are all accessible by the 
Atacama Large Millimeter/Submillimeter Array (ALMA).
The subsequent surveys of S2CLS have been carried out 
or are being carried out with the JCMT, as a large programme, 
on the COSMOS field \citep[e.g.,][]{Simpson2019},
as well as their follow up with ALMA \citep{Simpson2020}.

The North Ecliptic Pole (NEP) region
is one of the seven extragalactic fields 
included in the original S2CLS project \citep{Geach2017}.
Through a more recent JCMT large programme, 
which is one of the subsequent surveys of S2CLS, 
the original 850-$\mu$m area coverage 
has been extended from 0.6\,deg$^2$ to about 2\,deg$^2$,
allowing the detection of 549 submillimetre sources 
\citep[above 4$\sigma$,][]{Shim2020}.
Some counterpart identification of 
S2CLS-NEP 850-$\mu$m sources was already
presented in \citet{Seo2018}, 
using only a small subsample of sources 
with secure spectroscopic redshifts. 
Armed with the abundant multi-wavelength data sets, 
including the X-ray, optical, near-, mid-, to far-infrared 
(hereafter NIR, MIR and FIR, respectively)
and radio wavelengths 
\citep{White2010, Kim2012, Takagi2012, Krumpe2015, Nayyeri2018, Pearson2017, Pearson2019, Oi2021}, 
we now present the multi-wavelength counterpart identification 
of 850-$\mu$m sources located in the full NEP survey field 
\citep{Geach2017, Shim2020},
as well as their estimated physical properties.

This paper is organized as follows. 
In Section~\ref{sec:data}, 
we describe our 850-$\mu$m data
and the availability of ancillary multi-wavelength
observations.
We present the counterpart identification in
Section~\ref{sec:counterid}.
The models and parameters for SED fitting
and the related analysis are described in Section~\ref{sec:sedfit}.
Results based on the derived physical parameters,
such as the photometric redshift,
stellar mass, and SFR are
discussed in Section~\ref{sec:results}.
Our work is summarized in Section~\ref{sec:summary}.
Throughout the paper,
we use \textit{Planck}~2018 cosmological parameters 
for a flat $\Lambda$CDM model
\citep[][$\Omega_{\rm m, 0}=0.315$, $H_0=67.4\,$km\,s$^{-1}$\,Mpc$^{-1}$]{Planck2018},
and magnitudes are given in the AB magnitude system \citep{Oke1974}.

\section{Data}
\label{sec:data}

\begin{figure}
  \includegraphics[width=\columnwidth]{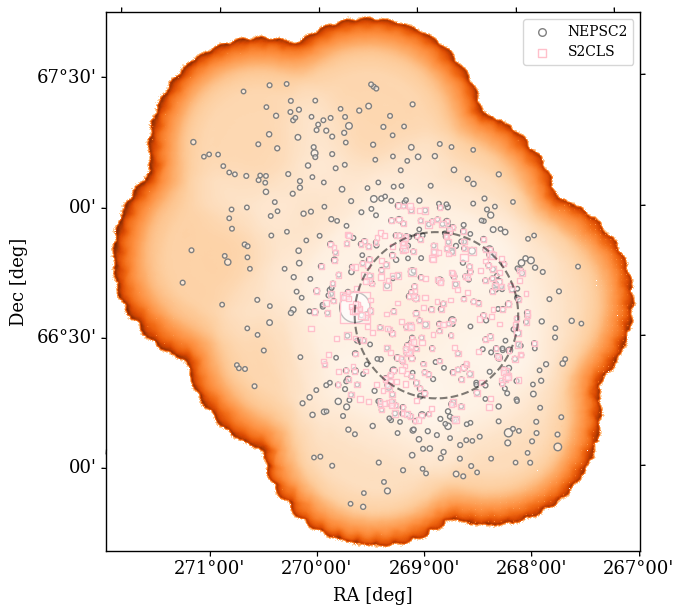}
        \caption{
	Submillimetre (850\,$\mu$m) 
	source positions overlaid on the variance map 
	from the NEPSC2$+$S2CLS image \citep{Shim2020}. 
	Open circles and squares indicate 850-$\mu$m sources 
	in the NEPSC2 catalogue 
	\citep[4$\sigma$ detection limit,][]{Shim2020}
	and the S2CLS catalogue
	\citep[3.5$\sigma$ detection,][]{Geach2017},
	respectively. Symbol sizes are scaled according to the 
	detection S/N. The dashed circle shows 
	the region that is called 
	the \textit{AKARI} NEP-`deep' field, 
	while the rest of the region  
	is included in the NEP-`wide' field. 
	For more detailed descriptions of the 
	\textit{AKARI} NEP fields,
	including other multi-wavelength data coverage,
	see \citet{Kim2021}.
        }
  \label{fig:sourceloc}
\end{figure}

\subsection{Submillimetre}
\label{sec:submm} 

The S2CLS project \citep{Geach2017} 
obtained 850-$\mu$m imaging over 0.6\,deg$^2$ 
centred on the North Ecliptic Pole, 
at an rms level of 1.2\,mJy beam$^{-1}$. 
The following North Ecliptic Pole SCUBA-2 survey 
\citep[NEPSC2,][]{Shim2020} 
expands the 850-$\mu$m area coverage to 2\,deg$^2$
at the rms level of 1--2\,mJy beam$^{-1}$.
The NEPSC2 source catalogue is constructed 
using an updated 850-$\mu$m mosaic map 
through mosaicking of the previously available S2CLS data 
and the newly obtained data
\citep[see][for more details]{Shim2020}.

For this work, 
we merge the two catalogues into one. 
The signal-to-noise ratio (S/N) cuts applied 
in the catalogue construction are
different in the two surveys. 
The NEPSC2 catalogue contains 549 sources with 
4$\sigma$ detections, 
while the S2CLS catalogue contains 330 sources 
with 3.5$\sigma$ detections.
Figure~\ref{fig:sourceloc} shows
the spatial distribution of 850-$\mu$m sources
from the two catalogues,
overlaid on the variance map over the NEP area
(bright colours represent areas with low variance). 
By merging the two catalogues, 
the number of `unique' 850-$\mu$m sources
over the NEP region is 647.
For overlapping sources, we use the information 
(coordinate and flux density) from the NEPSC2 catalogue. 
Note that one extremely bright submillimetre source
with S/N of 195 is Galactic
(the planetary nebula NGC~6543).
The possibly spurious sources located in a
petal-like pattern around it
are excluded in the merged 850-$\mu$m source catalogue.
The new catalogue 
is used as an input 
in the multi-wavelength counterparts identification.

\subsection{Multi-wavelength ancillary data}
\label{sec:multiwavelength}

\subsubsection{Radio}
\label{sec:radio}

The 20-cm continuum data 
(i.e., with an effective frequency of 1525\,MHz) 
of the NEP-deep field 
have been obtained using the Karl G. Jansky Very Large Array (VLA) 
during 2013--2014 (project ID: VLA/13B-361 and VLA/14A-469,
PI: K. Nakanishi), with a total integration time of about 9.5 hours. 
Data reduction to construct the primary beam-corrected 
image has been performed with standard methods using  
\textsc{casa}, the Common Astronomy Software Applications package
\citep[][]{CASAref}.
The final image has a size of a 0.35-degree radius circle, 
reaches an rms noise level of 6\,$\mu$Jy in the centre, 
and has a spatial resolution of around 1.5\,arcsec. 

We use the 20-cm source catalogue (as well as the image) in the 
radio identification of 850-$\mu$m sources. 
Source extraction is performed 
using the \textsc{blobcat} package \citep{Hales2012}
over the processed image,
along with the rms map created by 
\textsc{sextractor} \citep{Bertin1996}
as an additional input. 
The source detection threshold 
is set to $4.5\sigma$
to avoid detecting artefacts as sources 
while maximizing number of detected sources. 
The integrated surface brightness for each radio blob 
is defined as the 20-cm flux density 
for a radio object. 

Note that a Westerbork Synthesis Radio Telescope (WSRT) 
20-cm source catalogue is available over a 1.7\,deg$^2$ area 
on the NEP \citep{White2010}.
However, because of the relatively 
poor sensitivity (21\,$\mu$Jy rms)  
and spatial resolution (17.0\,arcsec$\,\times\,$15.5\,arcsec),
we do not use the WSRT source catalogue 
in the radio identification of 850-$\mu$m sources.

\subsubsection{Near-infrared to mid-infrared}
\label{sec:nirmir}

Several NIR photometric catalogues 
obtained from ground-based telescopes
are available over the NEP field: 
(1) CFHT/WIRCam $Y$, $J$, and $K_{\rm s}$ observations 
in the 47\,arcmin$\,\times\,$44\,arcmin region 
centred on the \textit{AKARI} NEP-deep field with 5$\sigma$ limiting 
magnitudes of 23.2, 22.8, and 22.5\,mag, respectively 
\citep{Oi2014};
(2) KPNO 2.1-m/FLAMINGOS $J$- and $K_{\rm s}$-band 
imaging
of the NEP-deep at 21.3 and 19.6\,mag 
\citep{Imai2007},
superseded by the previously described catalogue; 
(3) KPNO 2.1-m/FLAMINGOS $J$- and $H$-band
observations covering the entire NEP-wide field,
reaching down to a $5\sigma$ depth of 21.6\,mag 
in $J$ and 21.3\,mag in $H$ \citep{Jeon2014}; 
and (4) CFHT/WIRCam $K_{\rm s}$-band imaging observations
covering around 1\,deg$^2$ in the NEP-Wide 
to a $5\sigma$ depth of 21.9\,mag. 
The limited depth and the limited areal coverage
of these data sets
hamper the construction of 
an optical-to-NIR band-merged catalogue 
based on the NIR detection 
over the full 850-$\mu$m survey area. 
We thus rely on the near- to mid-IR 
source catalogues generated by space-based telescopes
for the identification of counterparts to
the 850-$\mu$m sources. 

The \textit{AKARI}/IRC NEP-deep and NEP-wide surveys 
make the NEP field a unique region of the sky 
where continuous MIR wavelength observations 
through 2--24\,$\mu$m are available.
The $5\sigma$ flux limits in the NEP-deep survey 
(covering approximately 2000\,arcmin$^2$) 
are 14.2, 11.0, 8.0, 48.9, 58.5, 70.9, 
117.0, 121.4, and 275.8\,$\mu$Jy 
for the $N2$, $N3$, $N4$, $S7$, $S9W$, $S11$, $L15$, $L18W$,
and $L24$-bands, respectively
\citep{Wada2008, Takagi2012}.
The number in the name of each band indicates 
the central wavelength
in $\mu$m.
In the 5.4\,deg$^2$ area of the NEP-wide survey, 
the $5\sigma$ flux limits are 
15.4, 13.3, 13.6, 58.6, 67.3, 93.8,
133.1, 120.2, and 274.4\,$\mu$Jy, respectively \citep{Kim2012}.
The full widths at half-maximum (FWHM) of the 
\textit{AKARI}/IRC point spread function (PSF) 
range over 4.5--7\,arcsec. 

Most of the \textit{AKARI} NEP-deep and NEP-wide survey
area was also covered by the \textit{Spitzer}/IRAC 
3.6- and 4.5-$\mu$m observations after 
the liquid helium run out, 
in Cycles~10 (Program ID: 10147, PI: J. Bock) 
and 14 (Program ID: 13153, PI: P. Capak).  
The data obtained during Cycle~10 
have $5\sigma$ sensitivities of 6.5 and 4.0\,$\mu$Jy 
in the 3.6- and 4.5-$\mu$m bands, respectively
\citep{Nayyeri2018}.
We create a new mosaic image using the data from both cycles 
and construct a source catalogue based on the new image.
In this catalogue, $5\sigma$ flux limit
is 1.6\,$\mu$Jy in both 3.6- and 4.5-$\mu$m bands. 
Considering the improved spatial resolution 
(with the PSF FWHM being 1.6\,arcsec) 
and the better sensitivity 
(a factor of 5 deeper at 4\,$\mu$m) 
of the \textit{Spitzer}/IRAC photometry 
compared to that of the \textit{AKARI}/IRC, 
the IRAC catalogue is mainly used
to identify short-wavelength counterparts of the 850-$\mu$m sources
based on colours (see Section~\ref{sec:colour-id}).

There is no large area survey using the \textit{Spitzer}/MIPS 
instrument over the NEP area. 
However, we search the \textit{Spitzer}
archive to find several small programmes located in the NEP 
that used the MIPS 24-$\mu$m band. 
For most cases, the depths are
more or less similar to that of the
\textit{AKARI}/IRC $L24$-band data from the NEP-deep/wide.
The PSF size is also similar (5.9\,arcsec FWHM) to that of
$L24$-band data. 
When available, the MIPS 24-$\mu$m
flux density is also included 
in the SED fitting of 850-$\mu$m source counterparts 
(Section~\ref{sec:sedfit}).

Regions around the ecliptic poles 
have been observed as a part of an all-sky survey by the
\textit{Wide-field Infrared Survey Explorer} (\textit{WISE}) 
satellite 
with a best sensitivity of 5--6\,mJy
at 22\,$\mu$m for $5\sigma$ detections
\citep{Cutri2012}. 
The \textit{WISE} photometric data points at 
3.4, 4.6, 12, and 22\,$\mu$m 
are included in the SED fitting as well, when available.

\subsubsection{Optical}
\label{sec:opt}

The most recent optical source catalogue over the entire 
4\,deg$^2$ of the NEP-wide survey area 
provides deep 5-band Subaru/HSC photometry, 
at $5\sigma$ limiting magnitudes of 
28.1, 26.8, 26.3, 25.5, and 25.0\,mag 
in the $g$, $r$, $i$, $z$, and $y$ bands, respectively \citep{Oi2021}. 
This supplements the previously available optical photometry data 
with the CFHT/MegaCam $ugriz$ bands \citep{Hwang2007}, 
Subaru/Suprime-cam $BVRiz$ bands \citep{Takagi2012}, 
Maidanak/SNUCAM $BRI$ bands \citep{Jeon2010}, 
and CFHT/MegaCam $u^*$ band \citep{Huang2020}. 

\subsubsection{Far-infrared}
\label{sec:fir}

Although the area coverage is limited to the NEP-deep area, 
the \textit{Herschel}/PACS 100- and 160-$\mu$m source catalogues 
are available with $1\sigma$ sensitivities of 
1.6 and 3.2\,mJy, respectively \citep{Pearson2019}.  
The entire NEP-wide area has been covered by the \textit{Herschel}/SPIRE
observations \citep{Pearson2017}.
While the detection sensitivities differ for source locations, 
the average $1\sigma$ rms 
at 250, 350, and 500\,$\mu$m is
9.0, 7.5, and 10.8\,mJy, respectively,
which is slightly above the SPIRE confusion limit
\citep{Nguyen2010}. The mosaicked \textit{Herschel}/PACS and SPIRE
images over the NEP as well as the blind and forced source catalogues 
are also provided by the \textit{Herschel} Extragalactic Legacy Project
\citep[HELP;][]{Shirley2021}. 
A more detailed description of the 
multi-wavelength catalogue compilation can be found 
in \citet{Kim2021}, along with the comparisons
between the survey areas 
over which different wavelength surveys are available.

\section{Counterpart Identification}
\label{sec:counterid}

\begin{figure}
  \includegraphics[width=\columnwidth]{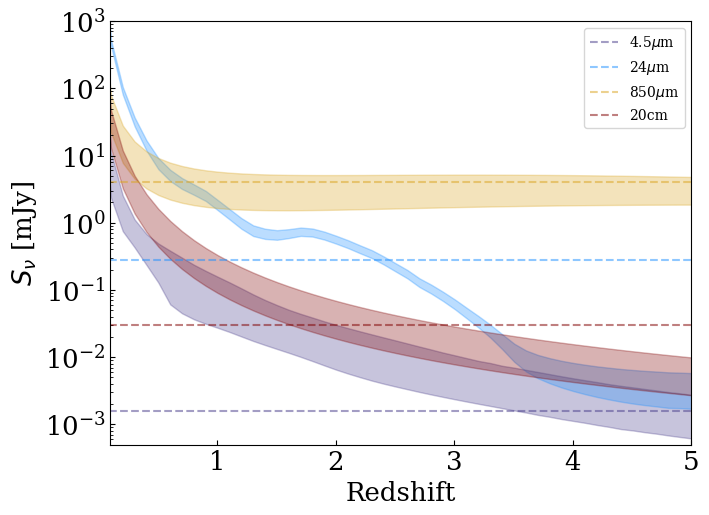}
        \caption{
	Expected flux density of an average SMG 
	\citep[from the ALESS survey;][]{daCunha2015}
	at different wavelengths 
	(4.5\,$\mu$m, 24\,$\mu$m, 850\,$\mu$m, and 20\,cm)
	redshifted to $z\,{=}\,$0--5. 
	The expected flux densities are illustrated by a range, 
	reflecting different spectral characteristics 
	in templates 
	obtained by averaging the faintest and the brightest
	objects in the SMG population. 
	Horizontal dashed lines are the $5\sigma$ flux limits 
    in the corresponding multi-wavelength ancillary data sets
    over the NEP field.
	In the case of 850-$\mu$m data, 
	we use the value of 4\,mJy,
	which is the 50 per cent completeness limit 
	for the NEPSC2 data.
        }
  \label{fig:fluxlim}
\end{figure}

Due to the large beam size
of the submillimetre single-dish telescope, 
the identification of counterparts 
at different wavelengths 
requires assumptions about the 
characteristics of 
galaxies
that can be selected in the submillimetre.
For instance, the use of radio wavelength information
is a reasonable approach to identify objects
responsible for submillimetre emission 
because of the well-known FIR-radio correlation 
of star-forming galaxies
\citep[e.g.,][]{Condon1992}.
In star-forming galaxies, 
the FIR emission is correlated with 
recently formed stars because 
it is produced by dust 
heated from the UV radiation of massive stars. 
The synchrotron radiation is also correlated with 
supernovae remnants from massive stars, 
therefore radio-detected galaxies are likely to be FIR bright. 
Mid-IR flux also correlates
with the total infrared luminosity and/or FIR
luminosity
\citep[e.g.,][]{Appleton2004, Takeuchi2005, Toba2017},
since both MIR and FIR emission
is generated from giant molecular clouds.
The actual relationship between the two is complex,
since the MIR polycyclic aromatic hydrocarbon
(PAH) features
are strongly affected by the contributions from an AGN,
but nevertheless, information at radio and/or MIR wavelengths 
can be used to identify counterparts 
for 850-$\mu$m sources.

Since the areal coverage of 20-cm data 
is only 0.38\,deg$^2$, smaller than the 2\,deg$^2$ of the 
NEPSC2 survey area, 
we need a method other than searching for radio emission 
to identify counterparts for submillimetre sources. 
As the spatial resolution of the MIR imaging data 
(7--24\,$\mu$m) is relatively poor,
using the shorter wavelength data (i.e., optical to NIR)
is a better strategy for counterpart identification. 
Previous studies such as \citet{Chen2016}
presented a counterpart identification method for 
single-dish 850\,$\mu$m sources 
based on the optical-to-IR colours. 
The method uses the assumption that 
submillimetre-selected populations are
heavily obscured galaxies
and/or are located at high redshifts,
having optical-to-IR colours that are distinguished 
from non-submillimetre populations. 
The method is quite successful, with the colour identification 
showing 80 per cent accuracy in identifying submillimetre source
counterparts that are identified in the higher-resolution 
interferometer follow-up data \citep{Chen2016}. 
We follow such a strategy to use colours
in order to identify counterparts of submillimetre sources 
in different wave bands.

Before we describe the counterpart identification process
in more detail, we discuss whether
the ancillary multi-wavelength data sets over the NEP
have enough sensitivity
to allow identification of an SMG population
over a wide range of redshifts.
Figure~\ref{fig:fluxlim} shows 
how the expected flux density in each filter 
depends on the redshift of an SMG. 
An averaged SED template from the 
ALESS SMG survey \citep{daCunha2015} 
is used here,
for which the total IR luminosity 
is scaled to $L_\mathrm{IR}=3.5\times10^{12}\,\mathrm{L}_\odot$.
The shaded flux density bands
show possible ranges
caused by the variation in the spectral shapes
among the SMG population with different total luminosity. 
Also plotted for comparison are $5\sigma$ flux limits 
in the ancillary multi-wavelength data sets 
over the NEP field: 
4.5-$\mu$m detection limits 
from the \textit{Spitzer}/IRAC mosaic; 
24-$\mu$m limits from the \textit{AKARI}/IRC 
NEP-wide field \citep{Kim2012}; 
and 20-cm limits from the VLA data. 
According to the plot, 
the available 24-$\mu$m ancillary data 
enable us to identify SMGs at $z<2.5$
if their SEDs are similar to those of the ALESS SMGs.
The radio data can be used for identifying $z<3$ SMGs. 
Finally, 4.5-$\mu$m images are 
deep enough for an identification of SMGs out to $z=5$,
although caution is needed
since the rest-frame optical-to-NIR 
SED shapes are more diverse 
than those of the rest-frame FIR SED, 
depending on star-formation history variations, for example.
Based on the comparison presented in Fig.~\ref{fig:fluxlim}, 
we use the VLA 20-cm and 
the \textit{Spitzer}/IRAC 3.6/4.5-$\mu$m data 
as a reference for identifying counterparts of 
submillimetre sources.

\begin{figure*}
  \includegraphics[width=0.95\textwidth]{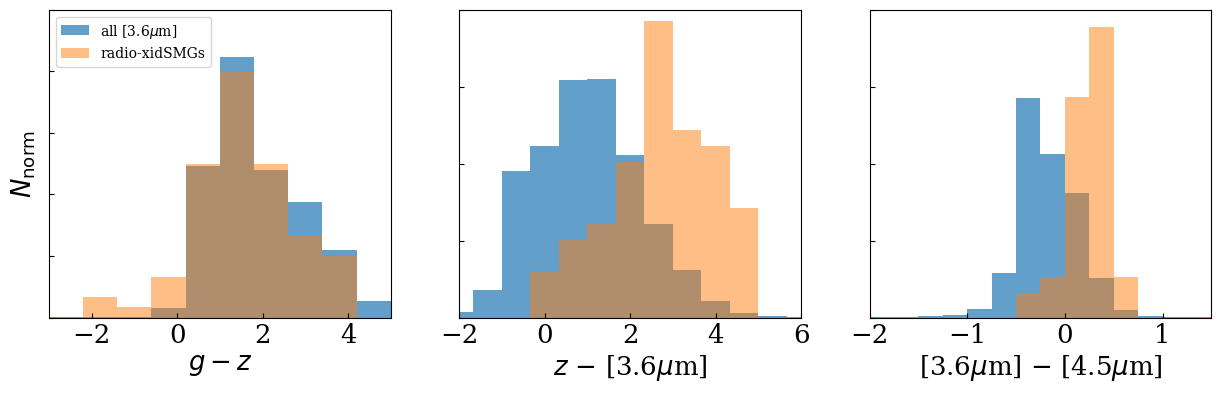}
        \caption{
        Colour distributions of all [3.6\,$\mu$m]-selected
        sources (i.e., background sources; blue) and
        radio-identified submillimetre source counterparts (orange).
        Histograms are normalized to unit area 
	to show the difference between two populations.
        }
  \label{fig:colourhist}
\end{figure*}

\subsection{Radio identification}
\label{sec:radio-id}

A conventional method of identifying counterparts 
at different wavelengths is 
to calculate the probability that the coincident match occurs 
within a specific radius, i.e., calculating $p$-values
based on the method outlined in \citet{Downes1986}. 
As used in many previous searches 
for single-dish SMG counterparts at different wavelengths
\citep[e.g.,][]{Ivison2007, Casey2013, Koprowski2016, Michalowski2017}, 
the $p$-value is defined as 
\begin{equation}
\label{eqn:pvalue}
	p = 1 - \mathrm{exp}\left(-\pi n \theta^2\right).
\end{equation}
\noindent Here $n$ is the source (surface) density 
and $\theta$ indicates the angular offset between 
the positions. 
In regions where the VLA imaging is available,
we cross-match our 850-$\mu$m source catalogue
with the radio source catalogue.
The radio source density in the NEP field is 
3200\,deg$^{-2}$, 
thus to get a false detection rate
less than 10 per cent,
we use a matching radius of 12\,arcsec.

The number of 850-$\mu$m sources 
that are located within the VLA field of view is 206
(note that the sensitivity of the VLA mosaic image
rapidly gets worse as we move to the outer edge).
Out of these 206 sources, 
we identify possible counterparts for 126 sources 
at radio wavelengths. 
Roughly speaking, this means that the radio identification rate 
is about 61 per cent. 
The identification rate is higher
for sources with larger 850-$\mu$m flux densities:
for example, the radio identification rate is
70 per cent for sources with $S_{850} > 5$\,mJy
and 32 per cent for sources with $S_{850} \le 5$\,mJy.
Most 850-$\mu$m sources have only one radio counterpart.
There are 10 sources with two radio objects  
within the matching radius, 
and two sources with three radio objects around the 850-$\mu$m position.  
The number of possible counterparts identified in the radio 
is 140 in total.
The offsets between the radio and 850-$\mu$m position 
are less than 8\,arcsec for 127 matches,
and 5\,arcsec for 108 matches.

It is difficult to estimate 
what fraction of submillimetre sources are actually
multiples of fainter objects \citep[e.g.,][]{Wang2011},
without the aid of submm/mm interferometric observations 
providing better spatial resolution. 
The only available such case for NEP submillimetre sources 
is one of the brightest sources ($S_{850}=32$\,mJy) 
that is observed at 1.1\,mm, 
which is found to be composed of two components 
\citep{Burgarella2019}.
Previous studies that performed higher spatial resolution
follow-up observations for single-dish selected sources
\citep[e.g.,][]{Hodge2013, Karim2013, Simpson2015, Simpson2020} 
have suggested that a significant (30--50\,per cent) 
fraction of submillimetre sources 
that are brighter than a few mJy 
at 850\,$\mu$m 
are composed of multiple components,
with an increasing fraction at higher flux densities. 
However, they have also showed that
most of the submillimetre flux density 
is contributed by the primary (i.e., the brightest) component, 
since secondary components are much fainter 
\citep[e.g.,][]{Michalowski2017, Hill2018, Simpson2020}. 
The fraction of 850-$\mu$m sources in our sample
with multiple radio counterparts 
among the radio-identified sources
is less than 10 per cent (i.e., 12/126).
Moreover, the average 850-$\mu$m flux density
of submillimetre sources with two or three 
radio counterparts 
is comparable to that of sources with only one 
radio counterpart.
If the radio emission 
is explained by the radio-FIR
correlation of star-forming galaxies, 
i.e., if the identified objects are not radio-loud AGN, 
the small fraction of multiple matches indicates
either that the actual multiplicity is relatively low 
at this flux density range ($S_{850}=3$--12\,mJy), 
or that the secondary counterparts are much fainter 
than the primary counterparts.

\begin{figure*}
  \includegraphics[width=0.95\columnwidth]{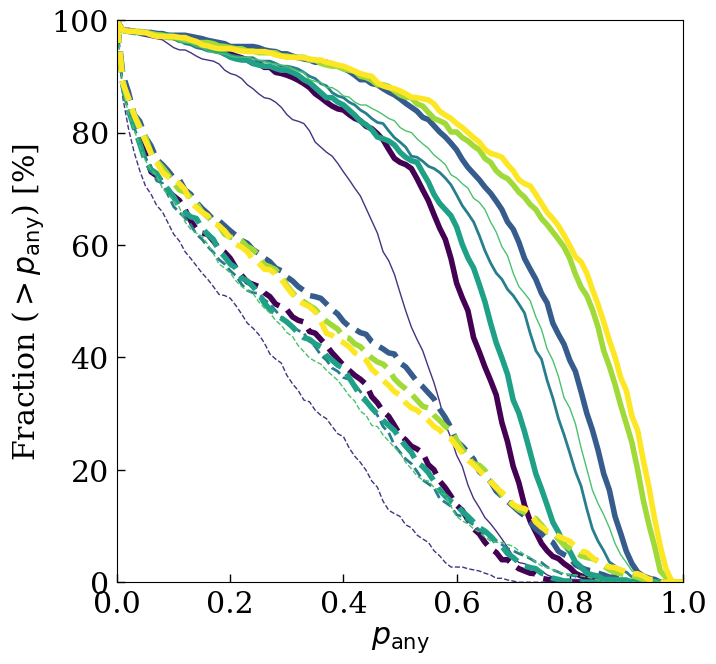}
  \hspace*{0.3in}
  \includegraphics[width=0.95\columnwidth]{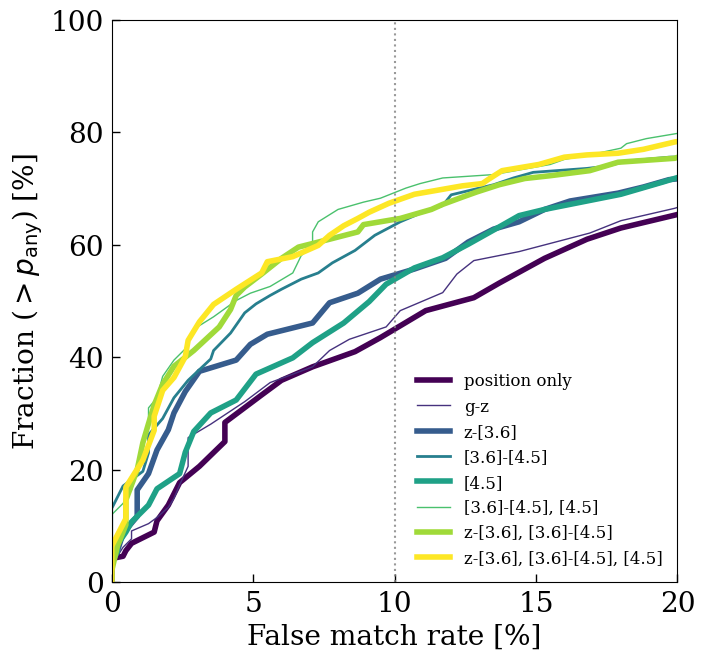}
  \caption{
        \textit{Left}: Cumulative fraction of submillimetre sources 
        for which counterparts are identified 
        with given $p_\mathrm{any}$ (Equation~\ref{eqn:pany}), 
        an output value from \textsc{nway}
	that indicates the probability of
	counterpart identification
	using different
	\textsc{nway} settings (solid lines).
        Dashed lines show the fraction of sources
        with counterpart identification as a function of
        $p_\mathrm{any}$ using a `fake-source' catalogue
	(i.e., using random positions). 
	To interpret, 
        solid lines show the completeness of a counterpart identification,
        and dashed lines show the false match rate.
        \textit{Right}:
        Comparison between the completeness and the false match rate.
        Note that the inclusion of the \textit{Spitzer}/IRAC bands 
	($[4.5\,\mu\mathrm{m}]$ magnitudes
	and $[3.6\,\mu\mathrm{m}]-[4.5\,\mu\mathrm{m}]$ colours)
        enhances the efficiency of a counterpart identification
	compared to the case using only position or position plus $g-z$ colour. 
	We choose a cut on $p_\mathrm{any}$ that matches to a
	false rate less than 10 per cent (dotted vertical line); this 
	corresponds to $p_\mathrm{any}>0.74$
	according to the left panel.
        }
  \label{fig:nway}
\end{figure*}

\subsection{Optical-to-NIR colour identification}
\label{sec:colour-id}

Although it is straightforward 
to identify counterparts at radio wavelengths, 
the source densities at shorter wavelengths 
(i.e., optical to MIR) 
are much higher than in the radio.
Thus we combine the positional offset with an additional prior
to calculate a probability for an object
to be a real counterpart to a submillimetre source.  
For this purpose, 
we use \textsc{nway} \citep{Salvato2018},
which is developed to identify
sources with large positional uncertainty
from data of better spatial resolution,
e.g., identifying optical counterparts 
for X-ray sources \citep{Hasinger2021}.
The \textsc{nway} code
searches possible matches to a source 
within a specified search radius
and calculates a probability for each object 
being a true counterpart 
based on the positional offset (i.e., a distance-based approach)
with a possibility of combining additional characteristics 
such as magnitudes or colours as priors.
Multiple priors can be tested simultaneously 
to find the most appropriate prior for source identification. 

Through a Bayesian approach, 
the probability that a source in a poor resolution image 
(850-$\mu$m source in this case) 
has {\it any\/} counterpart within the specified radius, 
$p_\mathrm{any}$, is calculated as 
\begin{equation}
\label{eqn:pany}
	 p_\mathrm{any} = 1-\frac{P(H_0|D)}{\sum\limits_i P(H_i|D)}.
\end{equation}
Here the prior $P(H)$ is 
the probability of a chance alignment 
of physically unrelated objects, 
and the posterior $P(H|D)$ 
can be calculated using information, obtained from 
the observed data $D$. 
$H_0$ indicates the no-counterpart hypothesis, 
while $H_i$ indicates a hypothesis that
the ${\it i\/}$-th object is a true counterpart. 
Based on this equation, 
large $p_\mathrm{any}$ values imply that 
at least one of the associations between different catalogues 
is likely to be physically related. 
The \textsc{nway} algorithm also provides 
relative posterior probabilities 
for every possible match ($p_i$ for the {\it i\/}-th object), 
by normalizing the total probability (except the non-matched case)
to be 1:
\begin{equation}
\label{eqn:pi}
	p_i = \frac{P(H_i|D)}{\sum\limits_{i>0}P(H_i|D)}.
\end{equation}
To find counterparts of the submillimetre sources 
(i.e., to find physically related matches between 
the catalogues), 
a probability cut for $p_\mathrm{any}$ 
should be applied to the results. 
This may not limit the counterpart identification 
to a single object, 
since it is possible that 
multiple objects show comparable $p_i$
to each other. 
Our adopted probability cut 
for identification of NEP submillimetre sources
is described in more detail below.

\begin{figure}
  \centering
  \includegraphics[width=\columnwidth]{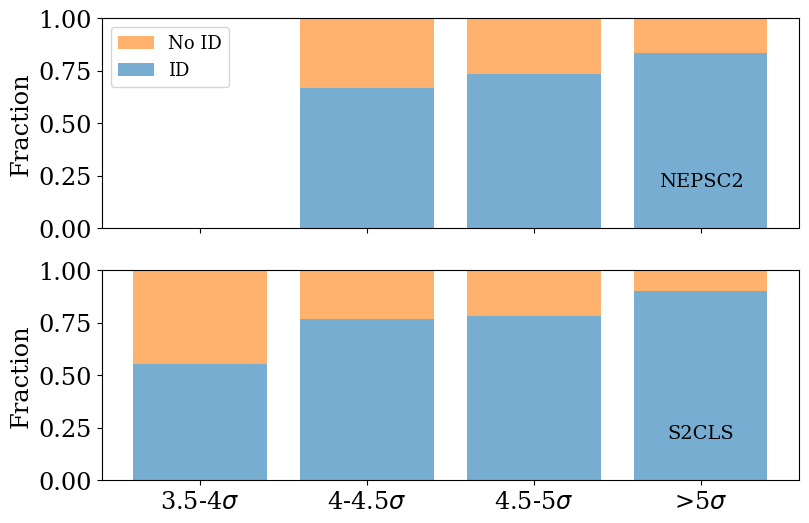}
  \caption{
	  Counterpart identification success rate 
	  based on the optical-to-NIR colours 
	  as a function of detection S/N at 850\,$\mu$m. 
	  The top and bottom panels
	  represent the identification rates 
	  for the NEPSC2 and the S2CLS catalogue sources,
	  respectively.
	  The fraction of 850-$\mu$m sources for which counterparts
	  are identified 
	  increases as the 850-$\mu$m S/N increases. 
  }
  \label{fig:idrate}
\end{figure}

Any attribute of the objects 
can be used as a prior. 
However, since the choice of prior affects 
the calculation of likelihood values through weighting factors, 
those properties for which the source population 
show different characteristics to others 
(i.e., background galaxies)
are good priors. 
To determine which attributes to use, 
we compare the colour distributions of
the counterparts for 850-$\mu$m sources 
that are identified in radio wavelengths
with that of other objects.
Figure~\ref{fig:colourhist} shows that
the radio-identified counterpart of submillimetre emission 
is clearly distinguished from other objects in 
$z-[3.6\mu\mathrm{m}]$ colours 
and $[3.6\mu\mathrm{m}]-[4.5\mu\mathrm{m}]$ colours,
while it is not the case in optical colours such as $g-z$.
Therefore, NIR colours can be used
to identify counterparts of the 850-$\mu$m emission 
in combination with the spatial offset. 
While any colours generated from the combinations of 
$z$ or $y$ with $[3.6\mu\mathrm{m}]$ or $[4.5\mu\mathrm{m}]$
are similarly useful for use as priors, 
we choose $z-[3.6\mu\mathrm{m}]$ 
because the $z$-band sensitivity is slightly better 
than that of the $y$ band. 

The left panel of Fig.~\ref{fig:nway} shows the
cumulative fraction of cases 
when counterparts are found 
with different probability cuts,
based on several different priors (colours and magnitudes). 
The completeness of identifications, 
i.e., the fraction of submillimetre-sources that do have counterparts, 
differs according to the probability cut. 
For example, 
in the position-only matching, 
adopting cases with $p_\mathrm{any}>0.7$
(in Equation~\ref{eqn:pany})
as robust identifications 
corresponds to the identification of 20 per cent of 
850-$\mu$m sources.
In order to determine a reasonable threshold 
for probability $p_\mathrm{any}$, 
we need an estimate of the false rate -- i.e., 
the probability that the identified counterpart 
is not a real counterpart but a coincident match. 
For this purpose, we create a `fake-source' catalogue 
by randomly changing the real 850-$\mu$m source positions 
and perform the counterpart identification process
with it using the same configuration.
The fractions of matches using a fake-source catalogue
are also illustrated in 
Fig.~\ref{fig:nway} (left panel) as dashed lines,
and the cumulative fraction in this case can be considered as 
the `false match' rate. 
A $p_\mathrm{any}$ cut for a false rate 
less than some specific value can be determined 
based on this plot. 

The right panel of Fig.~\ref{fig:nway} shows 
the relationship between the completeness ($y$-axis)
and the false match rate ($x$-axis) 
for different choices of prior. 
A parameter combination
that yields a high completeness and a low false rate simultaneously 
is the most efficient prior for 
counterpart identification. In our case,
the combination of 
$z-[3.6\mu\mathrm{m}]$ and $[3.6\mu\mathrm{m}]-[4.5\mu\mathrm{m}]$ colours with 
$[4.5\mu\mathrm{m}]$ magnitudes 
appears to be the best prior (yellow line in Fig.~\ref{fig:nway}).
The use of priors (colours and/or magnitudes)
results in a much better efficiency in counterpart identification 
than just using the nearest positional match
or the match using optical colour ($g-z$) and position.
High redshift galaxies and heavily attenuated galaxies are
`optically dark' 
\citep[e.g.,][]{Toba2020b},
and $z$-band magnitudes are not available for such galaxies. 
To account for this, we apply two
combinations of priors:
(1) $z-[3.6\mu\mathrm{m}]$, $[3.6\mu\mathrm{m}]-[4.5\mu\mathrm{m}]$
combined with $[4.5\mu\mathrm{m}]$;
and (2) $[3.6\mu\mathrm{m}]-[4.5\mu\mathrm{m}]$
combined with $[4.5\mu\mathrm{m}]$.

We apply a false rate cut of 10 per cent 
(i.e., $p_\mathrm{any}>0.74$) 
to define the identification as reliable. 
Note that a false rate of 10 per cent 
does {\it not\/} mean that 
the probability of a counterpart being false is 
10 per cent. Instead, it means that even a randomly defined
position would encounter a spatially close object 
with similar prior characteristics this fraction of the time.  
At this probability cut, the completeness reaches
around 65 per cent.
Among 647 sources in the 850-$\mu$m source catalogues,
counterparts of 643 sources are identified
based on the optical-NIR colours,
except for the four sources that are located
very close to saturated stars
or matched to local galaxies.
Then we have {\it reliable\/} identifications 
for 419 sources, corresponding to 65 per cent.
The number of sources with reliable identification
among the total number of submillimetre sources 
is defined as the identification success rate. 
Figure~\ref{fig:idrate} shows how 
the identification success rate 
changes as a function of detection S/N at 850\,$\mu$m. 
The results are shown separately for two different 
850-$\mu$m source catalogues, 
since the detection S/N for the same source 
is recorded differently in the two catalogues.
It is clear that the identification success rate
increases as the S/N increases. 
This is in line with the increasing false detection rate 
in the low S/N regime 
of the 850-$\mu$m source catalogue \citep{Shim2020}. 

We define any object with $p_i \ge 0.5$ 
(Equation~\ref{eqn:pi})
as a primary counterpart of the 850-$\mu$m source,  
since this means that the probability of that object
being the true counterpart 
is larger than the sum of probabilities of
all other objects within the search radius being counterparts. 
Objects with $0.2 \le p_i < 0.5$ are defined as
secondary counterparts.
Since the positional offset affects the probability 
calculation, we test if the difference between 
850-$\mu$m source coordinates 
measured in the NEPSC2 and S2CLS catalogues 
produces different results for finding most likely counterparts.
There exist only 16 cases that a primary counterpart
determined from the NEPSC2 position
is a secondary counterpart determined from the S2CLS position,
or vice versa, which accounts for 
only 7 per cent of the overlapping sources (16/222).
Therefore, the positional uncertainty at 850-$\mu$m
seems to have little effect on the counterpart identification.

Among the 419 submillimetre sources 
with reliable identifications using optical-NIR colours, 
40 sources have two (primary or secondary) counterparts, 
one source has three counterparts, 
while all other sources have only one primary counterpart. 
Again, this suggests that 
the probability of 850-$\mu$m sources 
being composed of multiple components
that are similarly contributing to 850-$\mu$m emission
is as low as 10 per cent (41/419).
Overall, counterpart identification
based on the optical-NIR colours
tends to select
a robust, single counterpart to each 850-$\mu$m source.

\subsection{Multi-wavelength photometry}
\label{sec:photom}

We construct the final counterparts catalogue 
by compiling all the counterpart identification results 
from Sections~\ref{sec:radio-id} and \ref{sec:colour-id}.
Among the 140 radio-identified counterparts, 
103 counterparts overlap with 
those identified in optical-NIR colours, although not 
all of them are considered to be reliable enough 
in terms of the 
$p_\mathrm{any}$ cut.
The overlapping cases are  
98 primary counterparts and five secondary counterparts.
The remaining 37 cases include: 
(1) four cases that are not detected shortward of 4.5\,$\mu$m
(i.e., 3.6$\mu$m-dropouts);
(2) 13 cases that are missed in the original IRAC source catalogue
because the object is not de-blended in the IRAC images,
though it is resolved in the optical images;
and (3) 20 cases for which other NIR-selected
objects are calculated to be more likely to be
the counterpart (i.e., the $p_i$ of the radio counterpart 
is calculated to be less than 0.2). 
Therefore, it can be said that 
the optical-NIR-colour identification 
recovers 84 per cent of the radio-identified objects (103/123).
This provides supporting evidence
for the reliability of optical-to-NIR
colour-based counterpart identification,
which is convenient, since 
there exists no large submm/mm interferometric
observation over this field to date.

\begin{figure*}
  \includegraphics[width=\textwidth]{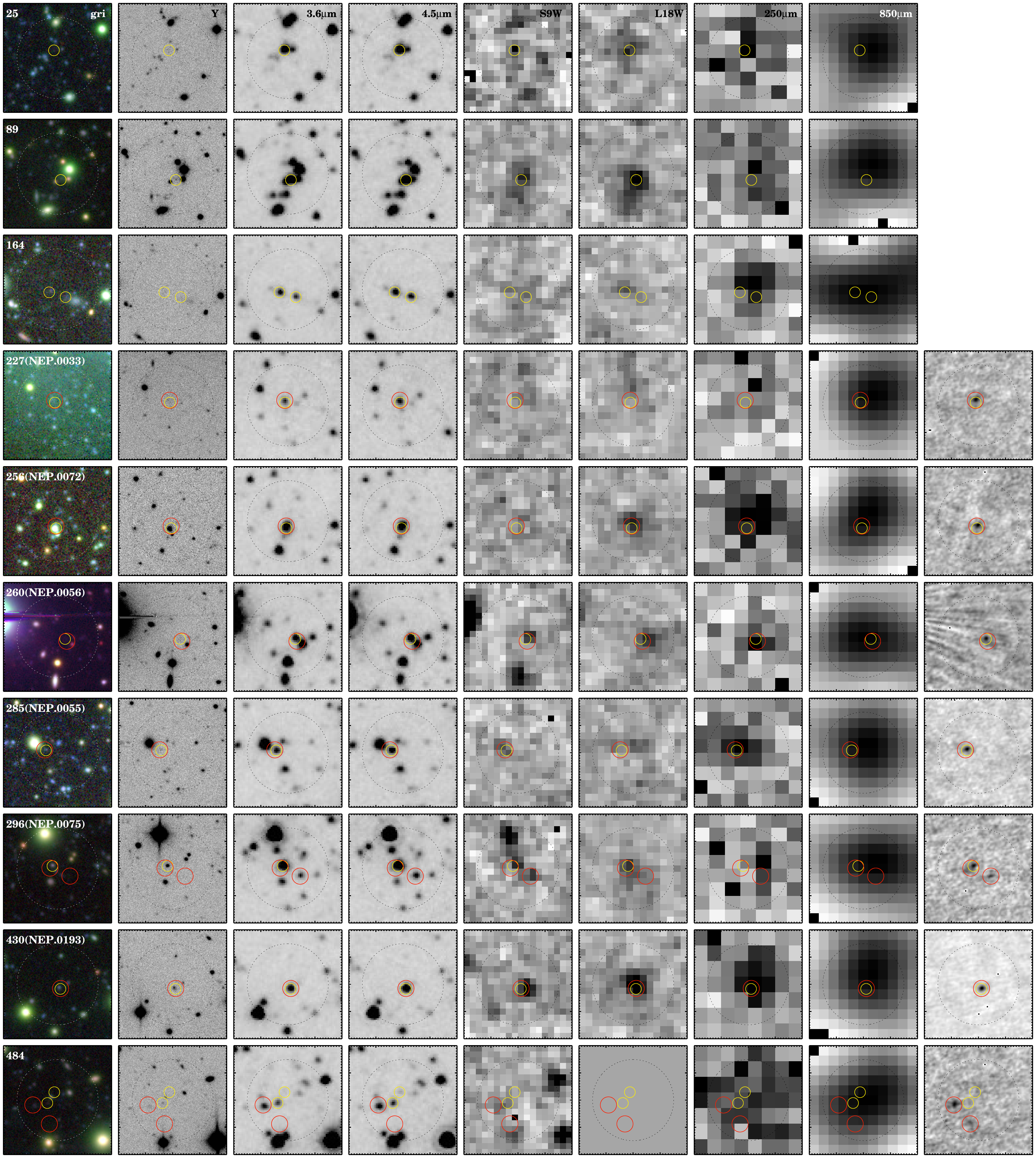}
  \caption{
          Examples of multi-wavelength
	  cut-out images of the 850-$\mu$m sources
	  and their counterpart identifications.
          All images have a size of 40\,arcsec $\times$ 40\,arcsec.
          Images listed are (from left to right):
          (1) an optical colour composite image
          (based on the Subaru/HSC $g$-, $r$-, and $i$-band
          images as blue, green, and red, respectively);
          (2) Subaru/HSC $y$-band image;
          (3) \textit{Spitzer}/IRAC 3.6-$\mu$m image;
          (4) \textit{Spitzer}/IRAC 4.5-$\mu$m image;
          (5) \textit{AKARI}/IRC 9-$\mu$m image;
          (6) \textit{AKARI}/IRC 18-$\mu$m image;
          (7) \textit{Herschel}/SPIRE 250-$\mu$m image;
          (8) JCMT/SCUBA-2 850-$\mu$m image;
          and (9) VLA 20-cm image.
          The large dotted circles represent
	  15 arcsec radius circles, 
	  which is twice the SCUBA-2 beam size. 
	  The small circles indicate the positions of counterparts
          identified in different wavebands, 
	  coloured red (3\,arcsec radius) in the radio, 
	  and yellow (2\,arcsec radius) at 4.5\,$\mu$m.
	  There exist cases that the counterparts are 
	  not detected at wavelength shorter than 3.6\,$\mu$m
	  (e.g., 227 and 260),
	  and cases where radio objects within the matching radius
	  are different from the most likely optical-NIR counterparts 
	  (e.g., 484). 
  }
  \label{fig:stamp}
\end{figure*}

\begin{table*}
	\caption{\textsc{CIGALE} parameters}
        \label{tab:cigale_fitparams}
        \begin{tabular}{ll}
                \hline
                Model and input parameters & Range \\
                \hline
                \hline
                Star-formation history: \texttt{sfh2exp}  \\
                \hline
                \hspace{4pt} e-folding time of the main stellar population model [Myr] & 1000, 3000, 5000 \\
                \hspace{4pt} e-folding time of the late starburst population model [Myr] & 30, 100, 9000 \\
                \hspace{4pt} Mass fraction of the late burst population &  0.0, 0.01, 0.1, 0.3, 0.5 \\
                \hspace{4pt} Age of the main stellar population in the galaxy [Myr] & 500, 1000, 3000, 5000, 10000 \\
                \hspace{4pt} Age of the late burst [Myr] & 50, 100, 300  \\
                \hline
                Stellar population: \texttt{bc03}  \\
                \hline
                \hspace{4pt} Initial mass function & Chabrier \\
                \hspace{4pt} Metallicity & 0.02 \\
                \hline
                Dust attenuation: \texttt{dustatt\_modified\_CF00}  \\
                \hline
                \hspace{4pt} Av\_ISM & 0.3, 0.6, 1.0, 1.6, 2.3, 3.0, 3.8, 5.0 \\
                \hspace{4pt} Power-law slope of the attenuation in the ISM & $-$0.7 \\
                \hspace{4pt} Power-law slope of the attenuation in the birth clouds & $-$1.3 \\
                \hline
                Dust emission: \texttt{dl2014}  \\
                \hline
                \hspace{4pt} Mass fraction of PAH &  0.47, 1.12, 2.50, 3.90 \\
                \hspace{4pt} Minimum radiation field & 1.0, 5.0, 10.0, 25.0, 40.0\\
                \hspace{4pt} Power-law slope index ${\tilde\alpha}$ ($dU/dM \propto U^{\tilde\alpha}$) & 2.0   \\
                \hspace{4pt} Fraction illuminated from $U_{\rm min}$ to $U_{\rm max}$ & 0.1 \\
                \hline
                AGN emission: \texttt{fritz2006}  \\
                \hline
                \hspace{4pt} Ratio of the maximum to minimum radii of the dust torus & 60 \\
                \hspace{4pt} Optical depth at 9.7\,$\mu$m & 1.0, 6.0 \\
                \hspace{4pt} Radial gas density gradient in the torus, $\beta$ & $-0.5$ \\
                \hspace{4pt} Angular gas density gradient in the torus, $\gamma$ & 4.0  \\
                \hspace{4pt} Full opening angle of the dust torus &  100.0  \\
                \hspace{4pt} Angle between the equatorial axis and line of sight & 0.001, 60.100, 89.990 \\
		\hspace{4pt} AGN fraction ($f_\mathrm{AGN}$) & 0.0, 0.1, 0.15, 0.2, 0.3, 0.4, 0.45, 0.5, 0.6, 0.7, 0.75, 0.8, 0.9 \\
                \hline
                Radio emission: \texttt{radio} \\
                \hline
                \hspace{4pt} FIR/radio correlation coefficient $q_\mathrm{FIR}$ & 1.5, 1.7, 2.0, 2.15, 2.3, 2.45, 2.58, 2.7, 2.8 \\
                \hspace{4pt} Power-law slope $\alpha$ of the synchrotron emission & 0.8 \\
                \hline
        \end{tabular}
\end{table*}

The full catalogue
of the identified counterparts of 850-$\mu$m sources
will be provided online
as supplementary material.
Table~\ref{tab:coord} shows several selected entries 
of the catalogue as an example.
The optical coordinates of the counterparts
are measured in the 
Subaru/HSC $g$-band images when available, 
and in the $y$-band images if the object is not detected 
in the $g$ band. 
If there is more than one radio counterpart, 
we add the suffix `a' to any radio object 
that is the closest to the 850-$\mu$m position 
and `b' and `c' in order of distance. 
The suffices `-1' and `-2' 
for the optical-to-NIR identifier  
indicate whether the object is a primary counterpart
or a secondary counterpart. 
Multiple secondary counterparts are marked as `-21', `-22', etc. 

Several examples of multi-wavelength cut-out images
around submillimetre sources
are shown in Figure~\ref{fig:stamp}, 
which are centred on the 
850-$\mu$m coordinates.
The positions of identified counterparts 
are also marked: red for radio identification, 
and yellow for optical-NIR-identification.
In the optical colour-composite images
and in the Subaru/HSC $y$-band images,
most counterparts of submillimetre sources 
are likely to be faint.
Optical-NIR colours of the counterparts 
are expected to be red 
from comparison of the $y$-band and 3.6-$\mu$m images.
Note that the coordinates of MIR emission ($>15\,\mu$m) 
with corresponding FIR positions
appear to be useful for identifying 850-$\mu$m sources
if the spatial resolution is provided,
as has been suggested by other FIR surveys 
\citep[e.g.,][]{Shirley2021}.

Multi-wavelength photometric data points for the 
identified counterparts 
are compiled through the nearest object match
using either the radio (when available) 
or the 4.5-$\mu$m coordinates. 
At optical wavelengths, 
the HSC $grizy$ catalogue \citep{Oi2021} 
is already combined with the IRAC 3.6- and 4.5-$\mu$m 
catalogue (initially using 0.7\,arcsec matching radius)
to be used in counterpart identification.
After visually checking cut-out images of 
each submillimetre source
in the $grizy$ bands, 
as well as in the 3.6- and 4.5-$\mu$m bands, 
we re-derive photometry for several objects 
that are missed in the HSC catalogue.
The reason why these are not included in the released catalogue 
is that while that catalogue is constructed  
using the photometry based on the $g$-band (deepest) detection, 
these objects are faint in the $g$ band.
We also revise the 3.6- and 4.5-$\mu$m photometric points
for some objects 
that are not deblended at the IRAC spatial resolution 
by the initial object detection process.
The $u$-band photometry, 
the CFHT $ugriz$, and $YJK$ photometric data points 
are also combined through the search in 
all the available ancillary catalogues
\citep{Hwang2007, Takagi2012, Oi2014, Huang2020} 
using a matching radius of 1\,arcsec. 
This matching radius is chosen to yield $p$-values
(see Equation~\ref{eqn:pvalue})
less than 0.05, considering the source number density. 
Additionally, 
the \textit{WISE} 3-, 4-, 12-, and 22-$\mu$m
photometric points are combined using a search radius of 
3\,arcsec, 
from the AllWISE data release \citep{Cutri2012}.
Considering the large FWHM of the \textit{WISE} data, 
we combine the \textit{WISE} points to only isolated objects, 
defined as objects that do not have neighbours 
of 4.5-$\mu$m magnitude difference less than 1\,mag 
within 3\,arcsec.

To mitigate possible source confusion 
caused by the poor spatial resolution, 
we measure MIR and FIR fluxes
of the identified counterparts 
using the object position
on the processed image, 
instead of finding matches in previous 
catalogues \citep{Kim2012, Takagi2012}.
The position we use 
corresponds to
either the radio coordinates 
(when available) or the coordinates measured at 4.5\,$\mu$m.
Flux densities at 2--24\,$\mu$m are measured  
in the \textit{AKARI}/IRC 2--24\,$\mu$m
and \textit{Spitzer}/MIPS 24-$\mu$m images,
utilizing the associated photometry mode of 
\textsc{sextractor} \citep{Bertin1996}.
The MIR fluxes are measured in 6-arcsec diameter 
apertures (which are used 
in deriving 3.6- and 4.5-$\mu$m fluxes),
then aperture corrections are applied.
In the \textit{AKARI}/IRC 2--4\,$\mu$m images of 
many objects, 
the fluxes derived from catalogue matching 
tend to be larger 
than the fluxes measured at the positions, 
due to source confusion. 
However at 7--24\,$\mu$m, 
fluxes measured with the two different methods 
(one matching to the blind detection catalogue, and
the other performing forced photometry at the given positions) 
are mostly comparable with each other. 
This suggests that the MIR flux is dominated 
by the most likely counterpart of submillimetre emission, 
and the contribution to the MIR flux 
by other optical/NIR objects
within the submillimetre beam size is not significant. 

At FIR wavelengths (100--500\,$\mu$m), 
we measure fluxes of the counterparts 
in the \textit{Herschel}/PACS 100- and 160-$\mu$m images, 
and in the \textit{Herschel}/SPIRE 250-, 350-, and 500-$\mu$m images 
that are provided by 
HELP \citep{Shirley2021}. 
Fluxes are measured using \textsc{xid+} \citep{Hurley2017},
a de-blender tool that 
applies counterpart positions as a prior.
Since the spatial coverage of PACS 100- and 160-$\mu$m 
images is small, SPIRE photometric points 
are essential to constrain FIR SED shape along with 
the 850-$\mu$m flux density. 
However, the FWHMs in \textit{Herschel}/SPIRE 250-, 350-, 
and 500-$\mu$m images 
are 17.6, 23.9, and 35.2\,arcsec, respectively, 
much larger than that of the SCUBA-2 850-$\mu$m image. 
Therefore even if we use 
the positional information of identified counterpart 
to deblend the SPIRE fluxes, 
there remain the chance of source confusion 
at the larger radius. 
Effects of such possibilities on the SED fitting
are described in more detail in Section~\ref{sec:source_confusion}. 

The number of photometric bands in which the flux density is available 
ranges from 3 to 36, 
with a median value of 11. 
We perform the SED fitting analysis (see the next section) 
to the identified counterparts
using all available photometric points.

\section{SED fitting}
\label{sec:sedfit}

\begin{figure*}
  \includegraphics[width=\textwidth]{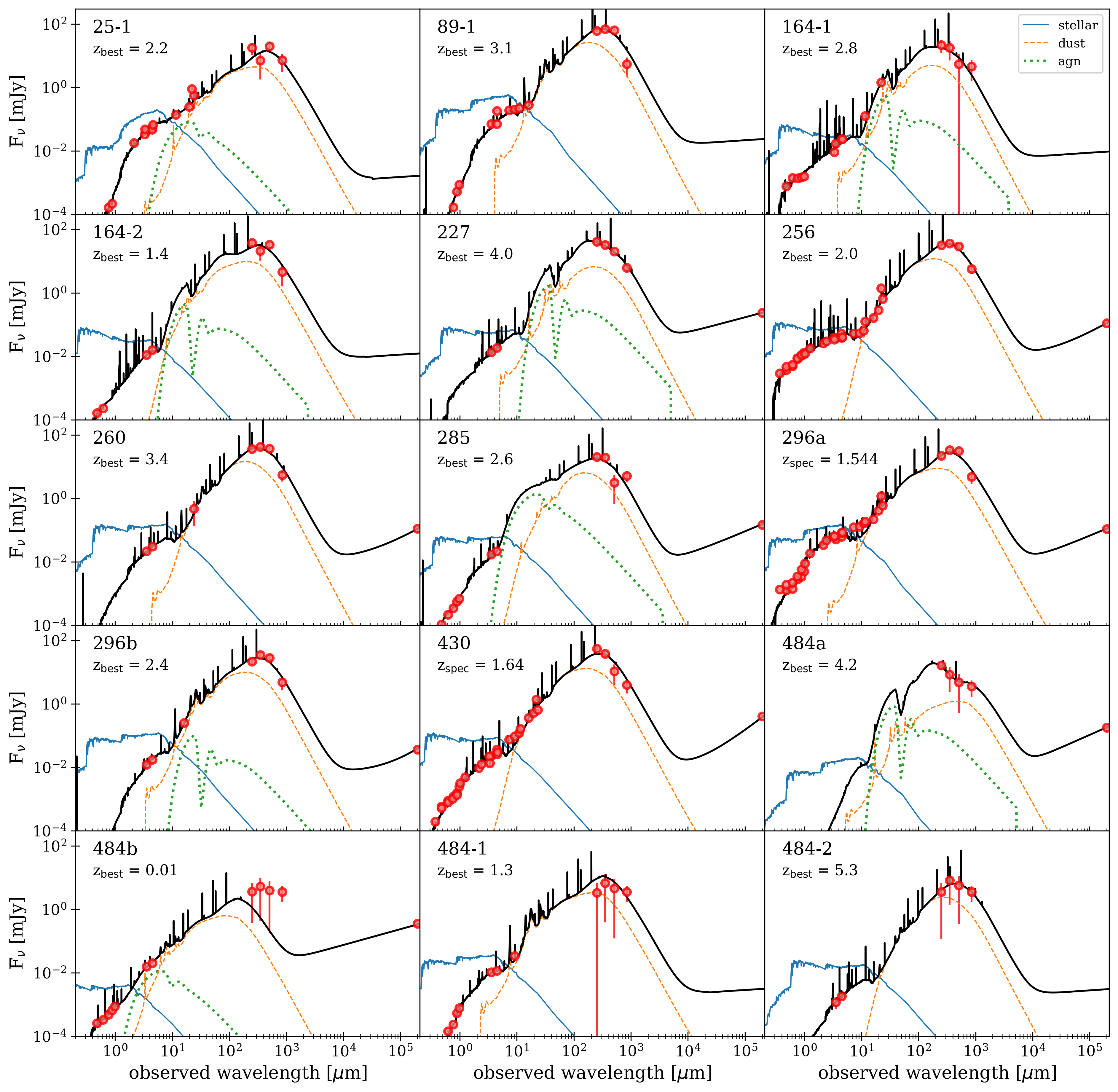}
  \caption{
	  Examples of the best-fitting SEDs
	  for the 850-$\mu$m source counterparts
	  from \textsc{cigale}.
          Each plot shows the best-fitting SED
	  for each object identified in Fig.~\ref{fig:stamp}. 
	  Black solid lines are model SEDs, 
	  which consist of the stellar component (blue lines, 
          before attenuation for clarity),
          dust emission component (orange dashed lines),
          AGN component (green dotted lines),
          and non-thermal radio emission at the longest wavelengths.
	  The number of photometric points 
	  used in the SED fitting is different for different objects,
	  with the maximum number being 36. 
          For objects with available spectroscopic
          redshifts (cases 296a and 430),
          the redshift is fixed during the SED fitting.
  }
  \label{fig:sedfit}
\end{figure*}

\begin{figure*}
	\includegraphics[width=\textwidth]{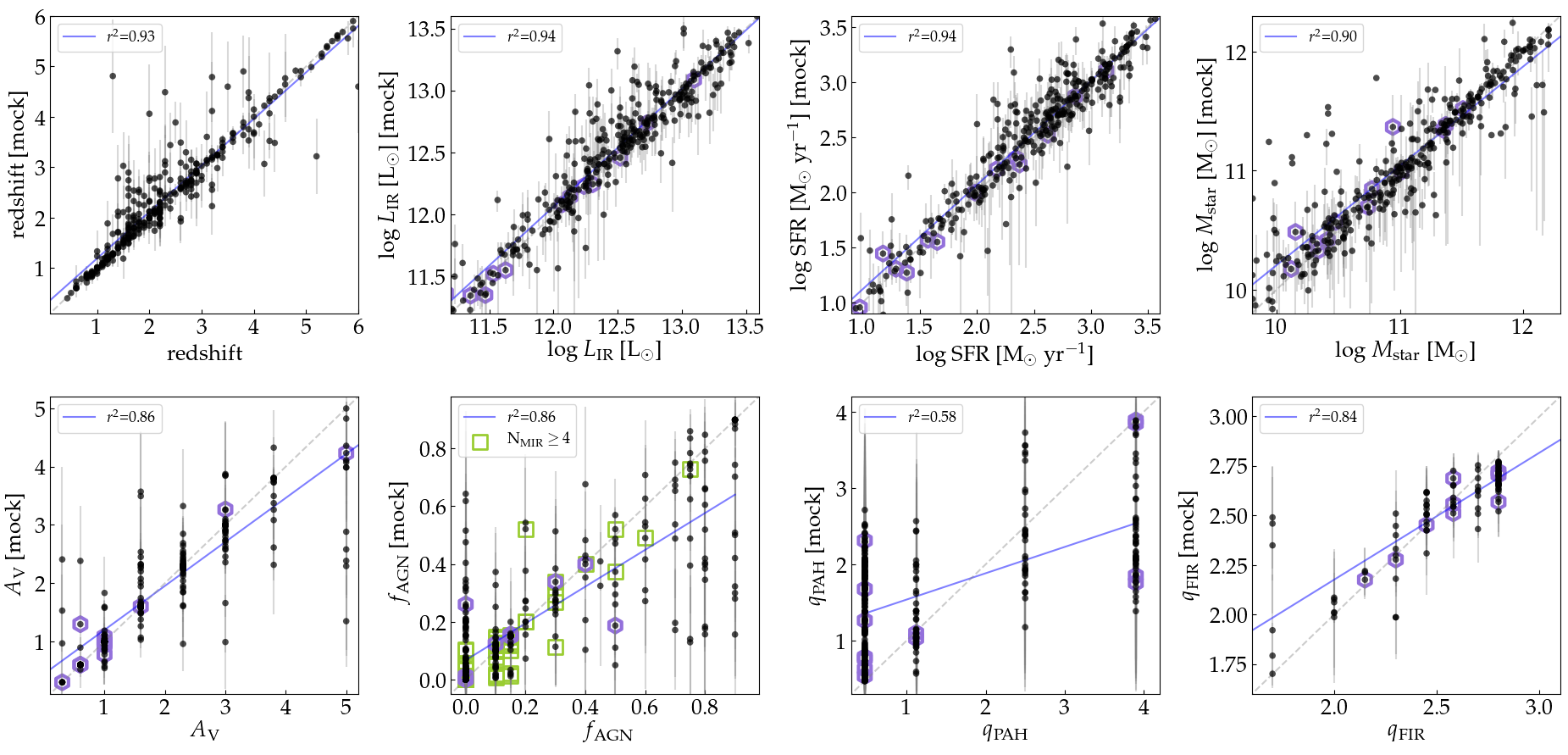}
	\caption{
	Mock catalogue tests to assess the quality of the 
	derived physical parameters. 
	The first seven panels represent the redshift, IR luminosity, 
	SFR, stellar mass, $V$-band dust attenuation $A_V$, 
	AGN fraction, and PAH mass fraction
	of the submillimetre sources, 
	while the last panel shows the FIR-to-radio 
	correlation coefficient $q_\mathrm{FIR}$
	for radio-identified submillimetre sources. 
	When the best-fitting SED is obtained for each object, 
	a mock catalogue is created using the best-fitting SED 
	and the observational flux uncertainties.
	This mock catalogue is again used as an input to the 
	SED-fitting procedure, using the same strategy, 
	then the estimated values (on the $y$-axis)
	are compared with the originally derived values 
	from the best-fitting SEDs ($x$-axis).
	Dashed lines show one-to-one correlations, 
	while the blue solid lines are the best-fitting linear 
	correlations between the $x$- and $y$-axis 
	with the indicated square of the correlation coefficients, $r^2$.
	Open hexagons are objects with spectroscopic redshifts.
	In the AGN fraction panel, objects with at least four points 
	in the 7--24\,$\mu$m wavelength range are marked as open squares.
	}
    \label{fig:mocktest}
\end{figure*}

\subsection{Fitting configuration}
\label{sec:sedfit_params}

Based on the multi-wavelength photometric catalogue
constructed in Section~\ref{sec:photom},
we derive physical properties of 
submillimetre source counterparts 
through SED fitting.
We use 
\textsc{cigale}\footnote{https://cigale.lam.fr/2020/06/29/version-2020-0/}
\citep[Code Investigating GAlaxy Emission,][]{Noll2009, Boquien2019},
which computes spectral models 
based on an energy balance principle
between the ultra-violet and the infrared wavelengths,
where the former mainly represents the energy from
direct stellar radiation
and the latter reflects emission reprocessed
by the dust.
By fitting the computed spectral models
to the observed photometric points, 
\textsc{cigale} allows us to investigate the 
physical properties of galaxies
including the star-formation rate (SFR),
dust luminosity, contribution of AGN
to the total dust luminosity (i.e., AGN fraction),
dust attenuation, 
and the masses of the stellar and dust components. 
For each output property,
the estimated value and its uncertainty
are calculated from 
the mean and standard deviation of the 
probability distribution function, 
which is weighted by the goodness of fit. 
Since we use the `photometric redshift'
mode of \textsc{cigale},
the photometric redshift (and its uncertainty)
is derived from the SED fitting as well.
The photometric redshift value is searched
in redshift ranges between 0.1 and 6.0,
with a redshift step size of 0.1. 

The models and input parameters we use
in running the modules of \textsc{cigale}
are summarized in Table~\ref{tab:cigale_fitparams}.
In order to determine which models to adopt
and what values to be used as inputs,
we refer to previous studies
on the FIR-selected galaxies \citep{Barrufet2020},
submillimetre-selected galaxies \citep{Seo2018},
and optically dark and/or red galaxies \citep{Toba2020b}.
Since it has been suggested that two-component star-formation histories
provide more accurate estimates
of the derived stellar masses of SMGs
\citep{Michalowski2014}, 
we use a double decreasing exponential
(\texttt{sfh2exp}) 
as the star-formation history. 
This consists of two stellar populations with different ages,
both described as bursts.
We choose the range of ages, e-folding timescales, and
late burst mass fractions to allow a wide range of
moderate and intense bursts.
By allowing a late burst mass fraction to be zero, 
the use of a double decreasing exponential 
may also test a classical single exponential. 
We use the stellar population synthesis model
of \citet{BC03}
with a \citet{Chabrier2003}
stellar initial mass function,
both of which have been widely adopted in 
previous studies of galaxy properties.
The metallicity is fixed to be solar,
since the SMGs are dust-rich galaxies.

For the attenuation of the stellar and nebular emission
(and estimation of the luminosity absorbed by dust),
we use the extended \citet{CF00} attenuation law, 
with wide parameter ranges 
based on the comparison between \citet{Calzetti2000}
and \citet{CF00} attenuation laws. 
With the \citet{CF00} approach, 
the attenuation levels of the young stellar population
(i.e., by the birth cloud) and the old stellar population
(i.e., by the ISM)
are treated separately, while they are combined to yield
the total attenuation.
\citet{Barrufet2020} have mentioned that
among different possibilities for dust attenuation laws,
the extended \citet{CF00} option
appears to best describe the SEDs of FIR-selected
galaxies. Likewise, in our analysis, the number of objects with 
better fits using the \citet{CF00} attenuation law 
is slightly larger than the number of objects
that prefer \citet{Calzetti2000}, 
which justifies the use of \citet{CF00}.
Note that for some objects, 
the use of one attenuation law produces a good-fitting result,
while the use of the other law is not at all successful. 
Nevertheless, we decline to apply different attenuation laws
to different objects 
since the use of different attenuation laws yields
different stellar mass estimates, 
such that the \texttt{dust\_modified\_CF00}
tends to result in estimates of the stellar mass
on average 20 per cent higher
compared to the use of
\texttt{dust\_modified\_starburst} for the same object.

To model dust emission through the MIR
to FIR by emission from 
PAHs, small grains and larger cold grains, the
\texttt{dl2014} dust templates are used
\citep{DL2007, DL2014}.
\citet{Seo2018} have mentioned that
the use of the \texttt{dl2014} module
generally results in better fits for SMGs
compared to the use of \texttt{dale2014},
since the former provides an extended variation of
the radiation field intensity and PAH mass fraction,
with an additional free parameter, namely the power-law slope.

The AGN models of \citet{Fritz2006}
are adopted, which consist
of three components: 
radiation from an active nucleus;
scattered emission by a dusty torus;
and thermal emission from the torus.
Most quantities (related to torus properties)
are fixed at default values,
while the AGN fraction is varied from 0 to 0.9.
The angle between the equatorial axis and
the line of sight is allowed to take the values 
0.001 (type-1 AGN), 89.990 (type-2 AGN), 
or 60.100 between the two. 
For objects with radio identification, 
we also include the radio module to model the synchrotron emission 
by varying the power-law spectral index $\alpha$ 
and the FIR-to-radio correlation coefficient $q_\mathrm{FIR}$
\citep[e.g.,][]{Helou1985, Yun2001}.
Since there is only one data point (20-cm) at radio wavelengths, 
the synchrotron power-law spectral slope $\alpha$
is fixed to 0.8, 
and the FIR-to-radio correlation coefficient 
is varied between 1.5 and 2.8.

\subsection{Fitting validation}
\label{sec:sedfit_valid}

Figure~\ref{fig:sedfit} shows 
the best-fitting SEDs of several selected counterparts. 
The diversity in the MIR regime 
of the SEDs 
suggests that a large number of photometric detections 
in this wavelength range 
is crucial in order to estimate the AGN contribution 
and for understanding the PAH emission. 
While the best-fitting SED is determined, 
the derived values for physical parameters 
from the \textsc{cigale} run 
are not the values
measured from the best-fitting SED.
Instead, they are evaluated by weighting all the models 
based on the fitting quality, 
with the best-fitting models having the largest weight. 
Probability distribution function for each physical parameter 
is derived by searching the maximum likelihood 
of the individual models located 
in given bins of the parameter space,
then the expected parameter value 
is calculated by taking the weighted sum of parameter space values
using the probability distribution function as weights 
\citep[see][~for details]{Noll2009}.

\begin{figure*}
        \includegraphics[width=\textwidth]{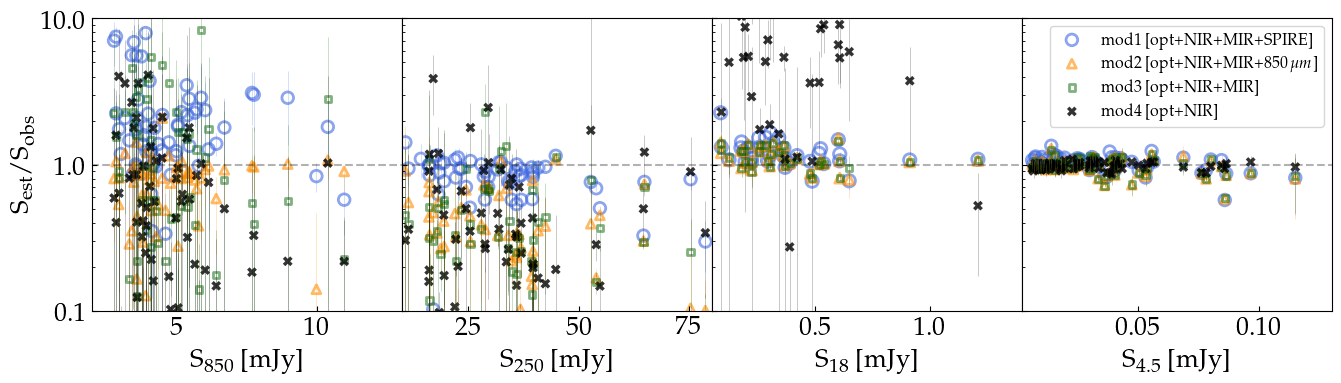}
        \caption{
	Ratios between the observed flux densities
        ($S_\mathrm{obs}$)
        and the flux densities estimated from the SED fitting
        ($S_\mathrm{est}$)
        at 850\,$\mu$m, 250\,$\mu$m, 18\,$\mu$m, and 4.5\,$\mu$m 
	    (from left to right),
        for different combinations of photometric points used in the SED fitting.
        Here `mod1' stands for the case
        where all photometric points other than the 850-$\mu$m point (i.e., SCUBA-2)
        are used in the SED fitting,
        while `mod2' represents the case
        where the SED fitting is done with flux densities in all available bands
        except the SPIRE 250--500-$\mu$m bands.
        Also compared are `mod3' and `mod4', which show the results from
        SED fitting using photometric points at wavelengths $\le24\mu$m,
	and $\le4.5\mu$m, respectively.
        }
        \label{fig:sconfusion}
\end{figure*}

The use of \textsc{cigale} for SED fitting provides an opportunity 
to estimate the reliability of 
the derived physical parameters.
Once the best-fitting SED is determined for each object, 
the flux densities in different filters 
can be calculated from the best-fitting SED. 
Since we know the exact `input values' of physical parameters, 
then by running \textsc{cigale} again 
using these calculated flux densities, 
along with the observed flux uncertainties, 
at the exactly same setting as in the original run, 
we can check whether the `output values' are 
consistent with the input values 
and see if the derived parameters are reliable. 
This is called a `mock test' in the \textsc{cigale} structure.

The results of the mock tests for our submillimetre sources 
are presented in Figure~\ref{fig:mocktest}.
The square of the correlation coefficient, $r^2$, 
is indicated for each physical parameter: 
(photometric) redshift; total IR luminosity; SFR; 
stellar mass; dust attenuation in the $V$-band; 
AGN contribution to the total IR luminosity; 
mass fraction of PAH molecules in dust mass; 
and FIR-to-radio correlation coefficient $q_\mathrm{FIR}$.
Since the redshift uncertainty affects the parameter values the most, 
for this purpose,
we use the 320 objects with $\Delta z/(1+z) \le 0.2$
(62 per cent of the entire 514 objects). 
In the cases of IR luminosity, SFR, and stellar mass, 
the input values and the output values are 
consistent within 0.1 dex, 
and the best-fitting linear correlation is quite close 
to the one-to-one line. 
This implies that the IR luminosity, SFR,
and stellar mass are reliable parameters
that can be determined through SED fitting.
The attenuation $A_V$, AGN fraction $f_\mathrm{AGN}$,
$q_\mathrm{PAH}$, and $q_\mathrm{FIR}$ 
are parameters with discrete input values. 
Nevertheless, these parameters also show $r^2$ values 
greater than 0.8, except the PAH mass fraction.
In the case of AGN fraction $f_\mathrm{AGN}$, 
defined as $L_\mathrm{AGN}/(L_\mathrm{AGN}+L_\mathrm{SF})$, 
the scatter appears to be larger 
than the aforementioned parameters, 
with $r^2$ being 0.86. 
The estimation of $f_\mathrm{AGN}$ in the \textsc{cigale} SED fitting
mostly relies on the warm dust emission 
from the putative AGN torus, 
with peak flux densities observed in the 7--24\,$\mu$m bands. 
This is supported by the fact that 
the correlation between the input and output values 
becomes tighter if the sample is limited to objects 
having at least four photometric points detected at MIR wavelengths
(see squares in the second panel 
of the bottom row of Fig.~\ref{fig:mocktest}).
The coefficient $r^2$
is even worse (0.58) for the PAH mass fraction $q_\mathrm{PAH}$,
although the true (input) $q_\mathrm{PAH}$ values
are still positively correlated
with the estimated $q_\mathrm{PAH}$ values by mock photometric points.
For radio-identified objects, the 
$r^2$ value for radio-to-FIR coefficient 
$q_\mathrm{FIR}$ is 0.84. 

The mock tests show that
we can trust IR luminosity, SFR, and stellar mass
for submillimetre counterparts if we limit the sample
to those having redshift dispersion $\Delta z/(1+z) \le 0.2$,
while caution is needed to use $A_V$ and $q_\mathrm{FIR}$ values.
It would be safe to use $f_\mathrm{AGN}$
only for classifying objects
into subgroups of objects
with high AGN contribution and low AGN contribution, 
since the quantitative $f_\mathrm{AGN}$ values
show too much scatter 
when there are not enough MIR photometric points available.

We also test if the use of a limited number of discrete values 
(such as step sizes for the redshift, $A_V$, AGN fraction, 
$q_\mathrm{PAH}$, and $q_\mathrm{FIR}$; 
see Table~\ref{tab:cigale_fitparams}) 
affects the correlation coefficient. 
Little difference is found between the $r^2$ values 
when we use half the step size presented in 
Table~\ref{tab:cigale_fitparams}. 
Rather, the `uncertainties' in the input values,
i.e., redshift and flux densities, 
are the main sources that increase the scatter. 
In the following analysis of physical parameters
(in Sections \ref{sec:sfrsmass} and \ref{sec:AGN}),
we mainly use objects with $\Delta z/(1+z) \le 0.2$, 
except for the case of the redshift distribution
(in Section \ref{sec:redshift}).
Additionally, the best-fitting SED for every object is 
visually inspected 
to see if the fitting quality is good enough.
This quality check for SED fitting 
is used to limit the sample in Section \ref{sec:results}. 

\subsection{Effect of different spatial resolutions to photometry}
\label{sec:source_confusion}

We mentioned that 
the flux enhancement due to a poor spatial resolution
appears to be small in the wavelength range 7--24\,$\mu$m, 
with the use of fixed position photometry 
and aperture corrections 
(Section~\ref{sec:photom}), 
which, however may not be the case for 250--500\,$\mu$m 
because of much larger FWHMs.  
In order to understand
how the large FWHM affects 
the flux densities at long wavelengths, 
we test the SED fitting procedure
using different combinations of photometric points,
i.e., without including flux(es) from certain band(s), 
then investigating whether the flux excluded 
is well reproduced 
by the SED fitting 
based on data from the other bands. 
Such a test strategy enables us to check for
consistency between photometric points from different bands. 
The different filter combinations used in the SED fitting are:
(1) optical to 500\,$\mu$m;
(2) optical to 24\,$\mu$m with 850\,$\mu$m added;
(3) optical to 24\,$\mu$m;
and (4) optical to 4.5\,$\mu$m.
To secure the reliability of counterpart identification,
objects used in this test 
are the radio-identified counterparts of 850-$\mu$m emission
that lie within 4\,arcsec at the 850\,$\mu$m position, 
except those that are multiples in 20-cm images. 

Figure \ref{fig:sconfusion} shows the results, i.e.,
the ratios between the `estimated' flux densities ($S_\mathrm{est}$)
and the `observed' flux densities ($S_\mathrm{obs}$),
in JCMT/SCUBA-2 850\,$\mu$m, \textit{Herschel}/SPIRE 250\,$\mu$m, 
\textit{AKARI}/IRC 18\,$\mu$m, and \textit{Spitzer}/IRAC 4.5\,$\mu$m data.
The estimated flux density is 
the average of all models weighted by the fitting quality.  
Thus it is close to the expected flux density 
in the given photometric band
inferred by photometric points in other photometric bands.
If the ratio is larger than 1, 
it means that the SED fitting overestimated the flux density 
in that filter, 
because the filters included in the SED fitting 
have enhanced flux densities compared to the 
filters not included in the SED fitting. 
The first panel of Fig.~\ref{fig:sconfusion} shows that 
850-$\mu$m flux densities are overestimated 
if the SPIRE 250/350/500\,$\mu$m fluxes 
are included in the SED fitting
while 850-$\mu$m flux is missing. 
The second panel shows the same trend that
the observed 250-$\mu$m flux density is higher 
than the estimation based on the 850-$\mu$m flux densities.
The observed 250-$\mu$m flux density is also high 
compared to the estimation using up to 
MIR photometric points ($\le24\,\mu$m),
with the average $S_\mathrm{est}/S_\mathrm{obs}$ of 0.6.
A general conclusion is that, 
while the source responsible for MIR emission 
is also responsible for 850-$\mu$m emission
(as the FWHM sizes in the MIR and 850\,$\mu$m are comparable), 
the flux densities at 250--500\,$\mu$m are possibly contaminated
by nearby fainter objects that are not identified as counterparts.

The third panel, for the 18-$\mu$m flux density, shows that 
if the SED fitting is done using only optical to NIR photometric points, 
i.e., bands with the highest spatial resolution, 
the observed MIR flux densities are not properly recovered. 
In most cases, the optical-to-NIR SED fitting suggests 
higher MIR flux densities than the observed values, 
not lower. 
Thus the difference cannot be explained 
by source confusion
in the MIR images with poor spatial resolution,
since the source confusion would increase the observed flux. 
Instead, this shows the limitation of properly recovering 
the complex MIR SED from the optical-to-NIR SED fitting. 
The NIR part of the SED is relatively well constrained, 
regardless of the inclusion of MIR and FIR data points,
again emphasizing the robustness of photometric redshifts, 
which are mainly driven by the optical-NIR photometric points
(last panel of Fig.~\ref{fig:sconfusion}).

From this analysis, we suggest that 
250--500-$\mu$m flux densities for submillimetre source counterparts
should be used with care, since there is the possibility
of flux enhancement
by the contribution of other fainter sources 
within the \textit{Herschel}/SPIRE beam. 
In other bands that have smaller or comparable FWHM as 850-$\mu$m image, 
such effect appears to be small.

\begin{figure}
	\centering
        \includegraphics[width=0.75\columnwidth]{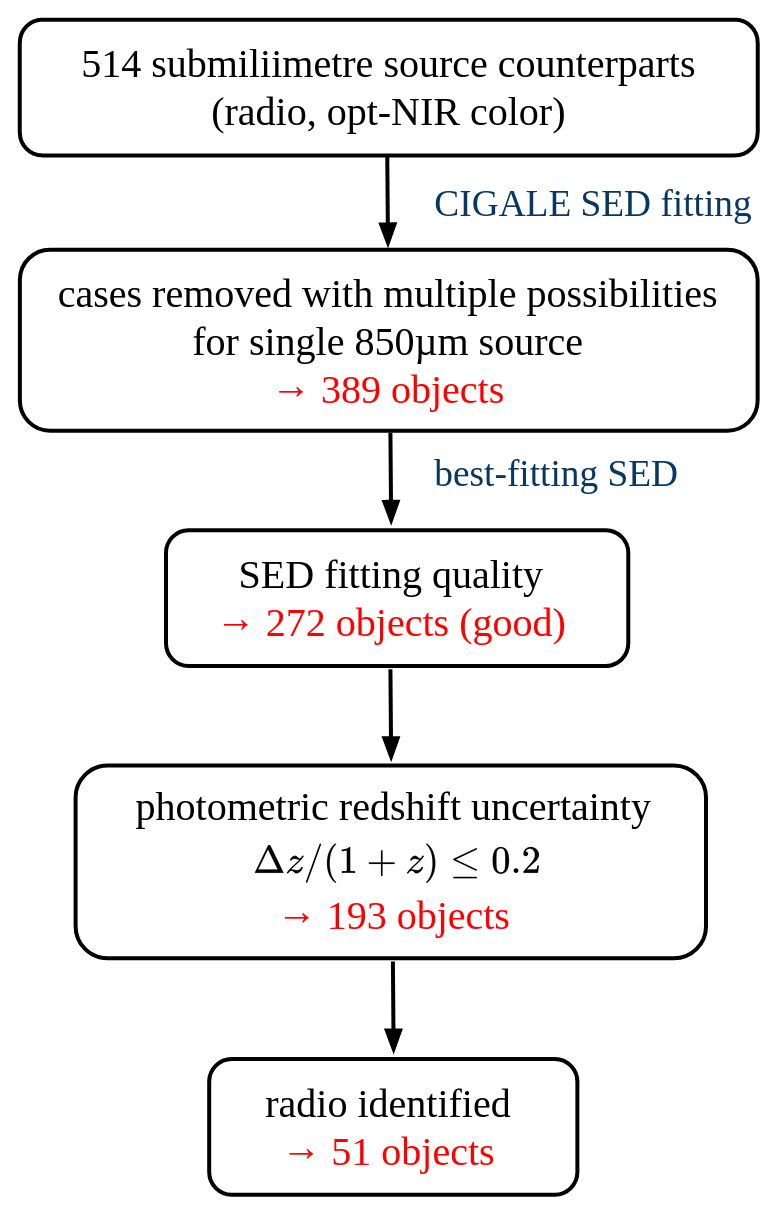}
  \caption{
	  Flow chart of sample definition process. 
	  Among 514 identified counterparts of submillimetre emission, 
	  we use 389 objects in the redshift distribution construction 
	  that are not likely to be contaminated by nearby objects. 
	  For SFR, stellar mass, and AGN contribution estimation, 
	  we use 272 objects with good SED fitting quality,
	  and discuss the effect of narrowing down the sample
	  by adding additional criteria for photometric redshift uncertainty.
	  }
  \label{fig:flow}
\end{figure}

\begin{figure}
        \centering
        \includegraphics[width=\columnwidth]{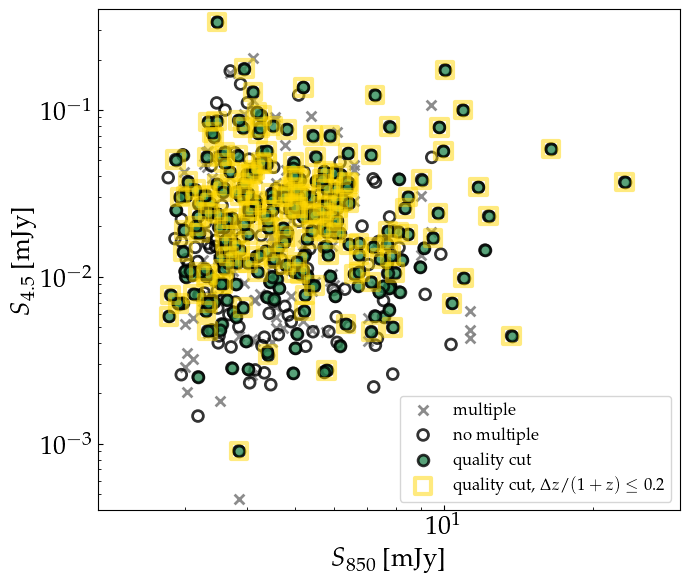}
  \caption{
	  Comparison between objects before and after
	  the sample definition process described in 
	  Fig.~\ref{fig:flow}. Crosses are 
	  submillimetre sources 
	  with more than one identified counterpart. 
	  Circles represent 389 objects that are left 
	  after excluding multiple identifications. 
	  Filled circles are 272 objects with good SED fitting quality,
	  by visually inspecting the best-fitting SED for 
	  each object. Among these, 193 objects enclosed with large squares
	  are objects with small uncertainty in photometric redshift 
	  estimation.
          }
  \label{fig:sample}
\end{figure}

\section{Results}
\label{sec:results}

Before presenting results on various physical parameters, 
we summarize the sample definition process with the flow chart 
in Figure~\ref{fig:flow}. Starting from 514 counterparts 
of 449 submillimetre sources, which are either identified 
by radio emission and/or by the combination of positional 
and optical-NIR colour information (we only include 
reliable cases for colour-based counterparts),
we perform \textsc{cigale} SED fitting for all objects. 
Then we exclude cases for which more than one counterpart 
is identified for a single 850-$\mu$m source,
since it is possible that 
flux densities of multiple counterparts are not properly
deblended in the images with large FWHMs. 
This leaves 389 objects (SMGs). 
Then by visually inspecting the best-fitting SEDs, 
we add tags to objects with good fit quality. 
These 272 objects comprise the basic sample 
that is used to draw conclusions about the nature of 
submillimetre sources. 
Our mock tests (Fig.~\ref{fig:mocktest}) showed that 
physical parameters are reliable for objects with 
small uncertainties in photometric redshifts, 
of which the number of objects is 193. 
Among these 193, 51 objects are radio-identified counterparts 
while the other 142 objects are optical-NIR colour identified. 

We investigate whether 
any bias can be introduced by the sample selection,
in terms of brightness in 850-$\mu$m and in near-infrared wavelengths.
Figure~\ref{fig:sample} shows the location of objects 
that satisfy the criteria in each step 
in the $S_\mathrm{4.5}$ versus $S_\mathrm{850}$ space.
850-$\mu$m sources
with more than one identified counterpart 
are slightly brighter in 850\,$\mu$m,
having $\langle S_\mathrm{850} \rangle=5.2$\,mJy, 
while the mean flux density is $\langle S_\mathrm{850} \rangle=4.8$\,mJy
for 389 objects if such multiple cases are removed. 
However, the difference is not large compared to typical 
flux-density errors at 850\,$\mu$m, 
and therefore reducing the sample size 
does not introduce strong biases 
in terms of 850-$\mu$m flux density of the submillimetre source population.
Moreover, the fraction of submillimetre sources having multiple counterparts 
does not increase as the 850-$\mu$m flux density increases. 
Since we do not have higher spatial resolution images 
at wavelengths close to 850\,$\mu$m, 
we can only rely on the counterpart identification strategy
introduced in this work, 
and conclude that the possibility of mis-identification of 
counterparts to bright submillimetre sources appears to be low. 
If we select objects with good SED-fitting quality, 
the mean 850-$\mu$m flux density is 
$\langle S_\mathrm{850} \rangle=5.2$\,mJy, again 
suggesting that the sample selection is not biased towards 
either higher or lower flux density sub-sample. 
In terms of 4.5-$\mu$m flux density, 
the $S_\mathrm{4.5}$ distributions are consistent 
with each other as the sample definition process proceeds
(with $\langle S_\mathrm{4.5} \rangle=0.027$--0.028\,mJy).
However, the photometric redshift uncertainty is affected 
by the S/N of the photometric points, 
thus $\langle S_\mathrm{4.5} \rangle$ is 0.035\,mJy,
if we consider 193 objects with good SED fitting quality 
and low photometric redshift uncertainty. 
The elimination of objects that have large $\Delta z/(1+z)$ 
removes optically-dark
counterparts of submillimetre sources 
that most likely have high photometric redshifts, $z>3$. 
In the following Sections~\ref{sec:sfrsmass} 
and \ref{sec:AGN}, we mark such objects with different symbols 
in the plots.

\begin{figure}
	\centering
        \includegraphics[width=\columnwidth]{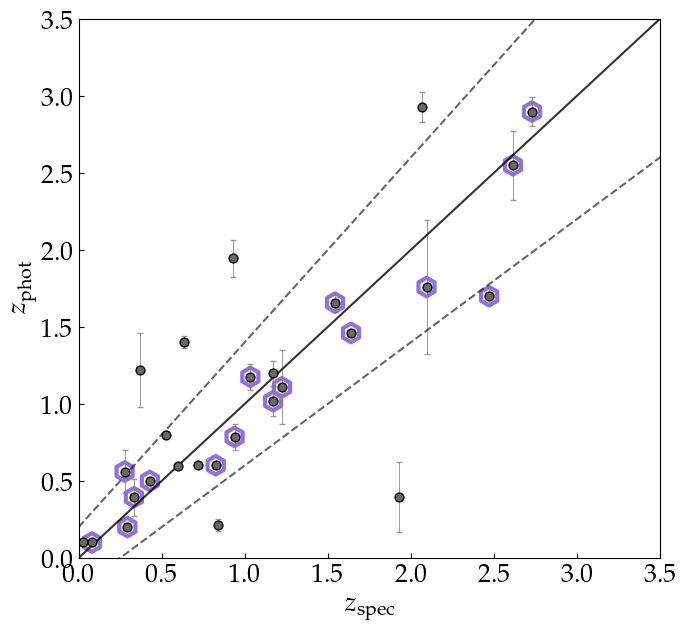}
  \caption{
        Comparison between spectroscopic redshifts
        (based on the optical or NIR spectroscopy)
        and photometric redshifts derived through
	\textsc{cigale} fitting (this work). 
        Objects enclosed by open hexagons have good
        spectroscopic redshift quality flags
	\citep[see][for details]{Shim2013}.
        The solid line indicates the $y=x$ relation
	and the dashed lines show ${\Delta z}/(1+z) = \pm0.2$. }
  \label{fig:compare_z}
\end{figure}

\subsection{Redshift and IR luminosity distribution}
\label{sec:redshift}

\begin{figure}
  \includegraphics[width=\columnwidth]{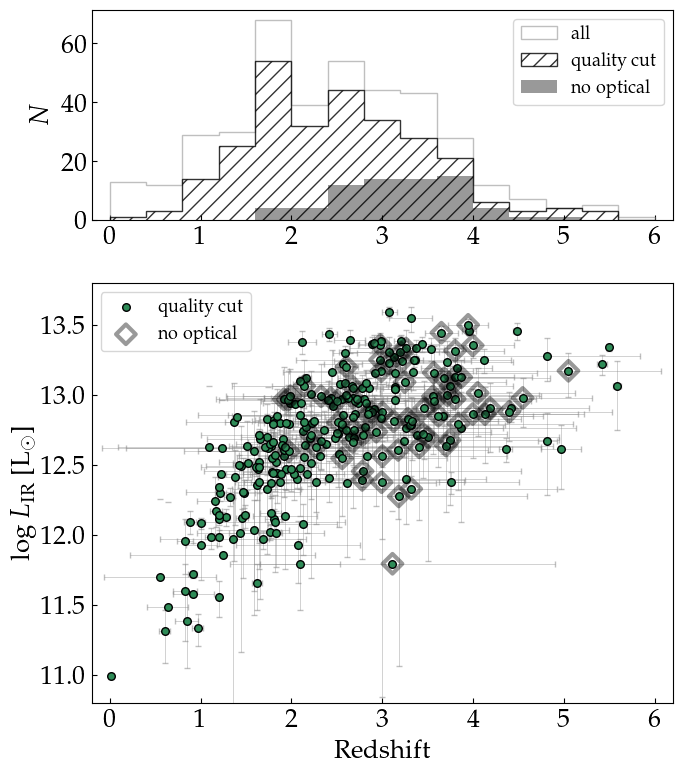}
  \caption{
	{\it Top}: Photometric redshift distribution of 
	submillimetre source counterparts.
	The faint histogram shows 389 objects 
	(after removing possible multiple counterparts), 
	while the hatched histogram shows 272 objects
	with good fitting quality.
	The mean value of the redshift distribution is
        $\langle z\rangle=2.5$ for the hatched distribution. 
	Also overplotted as a shaded histogram
        is the photometric redshift distribution 
	of a subgroup of submillimetre counterparts
	that are not detected in optical wavelengths, 
	with the mean value of $\langle z\rangle=3.2$.
	{\it Bottom}: Total infrared luminosity and the redshift 
	distribution of submillimetre counterparts. 
	While we expect that a few nearby and $z<2$ galaxies
	identified in the 850\,$\mu$m images
	are luminous infrared galaxies (LIRGs) with 
	$L_\mathrm{IR}>10^{11}\,\mathrm{L}_\odot$, 
	most submillimetre sources have IR luminosity larger than 
	$10^{12}\,\mathrm{L}_\odot$, 
	comparable to that of local ULIRGs. 
  }
  \label{fig:zhist}
\end{figure}

There are several spectroscopic redshift catalogues available 
over the NEP field \citep[e.g.,][and other unpublished 
redshift lists]{Shim2013},
from observations using 
MMT/Hectospec, WIYN/Hydra, Keck/DEIMOS, and GTC/Osiris, 
in addition to objects with spectroscopic redshifts 
that can be found from the 
NASA Extragalactic Database. 
We search these catalogues for matches to 
submillimetre source counterparts 
using coordinates constrained in the
radio or 4.5\,$\mu$m.
Most of the spectroscopic follow-up survey programmes 
over the NEP field targeted 
MIR-selected sources at optical 
and/or NIR wavelengths, and thus 
are likely to be limited to NIR-bright objects.
In total, 27 counterparts are matched to objects
in the spectroscopic redshift catalogues 
within a radius of 1\,arcsec. 
However, spectroscopic redshifts for 11 objects are 
assigned low quality flags, 
corresponding to cases where only one emission line
is used to estimate the redshift 
or the redshift is highly unreliable due to the poor
S/N of the spectrum \citep[see][for details]{Shim2013}. 

\begin{figure*}
    \includegraphics[width=0.99\textwidth]{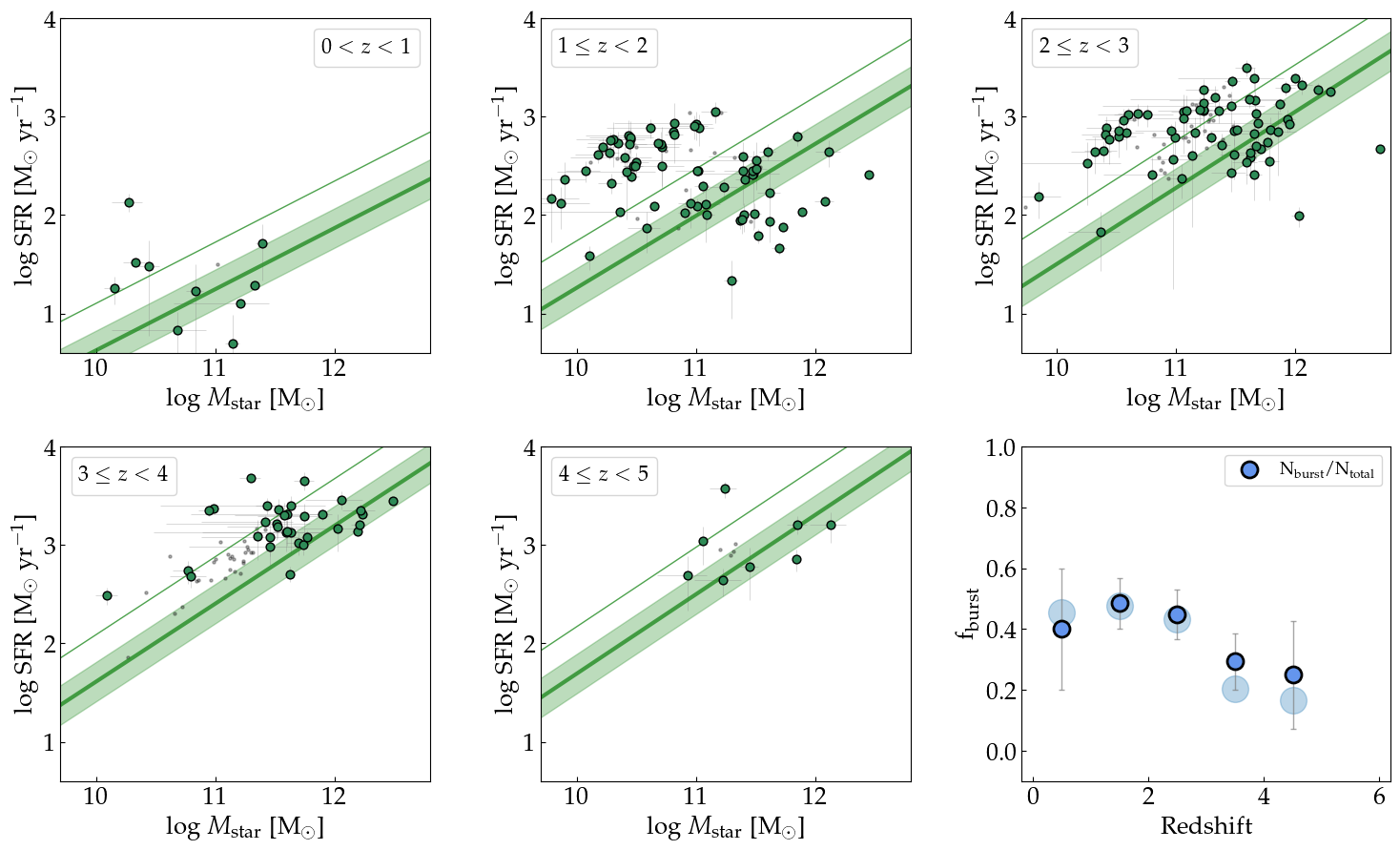}
    \caption{
            Star-formation rate versus stellar mass
            of submillimetre source counterparts
            in different redshift bins.
            Filled circles are objects with
	    $\Delta z/(1+z)\le0.2$,
	    while small dots are objects with larger uncertainty 
	    in the derived photometric redshift.
            The main sequence of star-forming galaxies from
            \citet{Speagle2014} is plotted
            with thick lines at redshifts $z=0.5, 1.5, 2.5, 3.5,$ and $4.5$.
            Shaded regions indicate $\pm0.2$\,dex from the
            star-forming main sequence.
            Thin lines show the boundary where the
            SFR of galaxies is 3 times that of the
            main sequence galaxies with the corresponding stellar mass.
            Objects above the thin lines are classified as
            `bursting' star-forming galaxies.
            In the bottom right panel, we show
            the number fraction of objects that are classified as
            bursting galaxies as a function of redshift.
    }
    \label{fig:sfrsmass}
\end{figure*}

Figure~\ref{fig:compare_z} shows a
comparison between the photometric redshifts
derived from \textsc{cigale} SED fitting 
and the spectroscopic redshifts. 
Objects with a good spectroscopic redshift quality flag,
i.e., a secure redshift identified with several
emission lines, are indicated 
with open hexagons on the plot.
For such objects, 
the derived photometric redshift 
is consistent with the spectroscopic redshift 
within ${\Delta z}/(1+z)$ of 0.2. 
Although the number is limited, 
the consistency between the photometric redshifts
and the spectroscopic redshifts
supports the reliability of the photometric redshifts
of identified SMGs.

The distribution of photometric redshift 
of SMGs 
is presented in the top panel of Fig.~\ref{fig:zhist}.
The faint grey histogram shows 
the redshift distribution of 389 objects
that are not among the multiple identified counterparts 
to a single 850-$\mu$m source, 
while the hatched histogram shows the redshift distribution of
272 objects with good SED fitting results through a visual check. 
The mean of the photometric redshift distribution for 
all 389 objects is $\langle z\rangle=2.4$
and for the 272 objects with good fitting quality, it is
$\langle z\rangle=2.5$, 
both with the standard deviation of 1.0.
Our value, $\langle z\rangle=2.5$,
is consistent with the redshift distribution of
the 850-$\mu$m selected SMGs
in earlier studies \citep{Chapman2003, Chapman2005},
and in other cosmological survey fields
\citep[e.g.,][]{Chen2016, Michalowski2017, Zavala2018, Dudzeviciute2020}.
There are cases where
the identified counterpart is not detected at wavelengths 
shorter than the NIR (i.e., $\le3.6\,\mu$m), 
or even at wavelengths shorter than the MIR.
Possible reasons for these non-detections include
heavy dust attenuation, high-redshift,
or a combination of both.
69 objects that are optically undetected show 
good fitting quality, 
and the mean redshift of this subgroup of objects 
is $\langle z\rangle=3.2$, 
higher than the mean redshift of the entire sample.

Figure~\ref{fig:zhist} (bottom panel) shows 
the total IR luminosity distribution of 
850-$\mu$m source counterparts as a function of redshift. 
As expected from the effect of the negative $K$-correction 
at 850\,$\mu$m (see Fig.~\ref{fig:fluxlim}), 
the lowest IR luminosity range 
that can be probed is 
consistently of the order 
$10^{12}\,\mathrm{L}_\odot$
over the redshift range $2<z<4$. 
The IR luminosity range of the sources
is $10^{11}$ to $3\times10^{13}\,\mathrm{L}_\odot$.
The optically undetected objects
that are identified as counterparts to 850-$\mu$m sources
are not the systems with the largest IR luminosities; 
instead, they are likely to be more heavily attenuated 
galaxies, as found by \citet{Toba2020b}.
No clear trend is found
between the 850-$\mu$m flux density ($S_{850}$)
and the photometric redshift.
There are several objects with $z_\mathrm{phot}>5$, 
which could be rare high-redshift dusty star-forming galaxies.  
These candidate $z_\mathrm{phot}>5$ objects are worth
further investigation through follow-up observations.

\subsection{SFR and stellar mass}
\label{sec:sfrsmass}

The SFR and stellar mass are physical quantities
that can be constrained more robustly by SED fitting
compared to other parameters (Section~\ref{sec:sedfit_valid}). 
The mean SFR of the 514 identified counterparts 
(when any sample selection cut is not applied)
is $\langle\mathrm{SFR}\rangle=220$\,M$_\odot$\,yr$^{-1}$,
and the mean stellar mass is 
$\langle M_\mathrm{star}\rangle=7.9\times10^{10}$\,M$_\odot$.
The SFR and stellar mass values are similar 
in magnitude to those derived for 
$z\simeq2$ submillimetre-selected galaxies 
in previous studies
\citep[e.g.,][]{Chen2016, Michalowski2017, Dudzeviciute2020}.
These two parameters are used to 
understand the nature of a galaxy, i.e.,
whether it is in a stage of quiet, passive evolution
or in a stage of vigorous star formation. 
Since the observationally identified 
main sequence of star-forming galaxies 
depends on redshift \citep[e.g.,][]{Speagle2014, Nelson2015}, 
we divide submillimetre source counterparts 
into redshift bins.

Figure~\ref{fig:sfrsmass} shows 
the relationship between the SFR and the stellar mass 
for the 272 
submillimetre source counterparts 
at different redshifts (with good SED fitting quality), 
compared to the main sequence of star-forming galaxies
in the corresponding redshift bin
\citep[][marked as lines]{Speagle2014}.
Objects with large photometric redshift uncertainty 
are marked as small dots, 
while objects with relatively secure photometric redshifts
are marked as filled circles.
At redshift bins $2\le z < 3$ and $3\le z < 4$, 
submillimetre sources with larger stellar masses show higher SFRs, 
following the trend of the star-forming main sequence, 
while most objects are located above the main sequence. 
At $1\le z < 2$, on the other hand, the scatter in the SFR for the 
given stellar mass is relatively large 
compared to the cases of higher redshift. 
Massive ($\mathrm{log}_{10}(M_\mathrm{star}/\mathrm{M}_\odot)\ge11.3$) 
objects at this redshift bin 
show lower SFR than that of main sequence galaxies.
Their SEDs suggest that they harbor non-negligible old stellar population 
reflected by a distinguished 4000\,\AA\, break
in addition to the dust component, 
which result higher stellar mass and low SFR. 
Note that objects of which SED fitting quality is 
not good enough are excluded in the SFR-stellar mass diagram. 
Since the quality of SED fitting is affected 
by the S/N in the flux densities, 
it is expected that the excluded objects are fainter, 
dominating the lower stellar mass range. 
The mean stellar mass of objects with good and bad 
SED fitting quality is
$\langle M_\mathrm{star}\rangle=1.6\times10^{11}$\,M$_\odot$
and $\langle M_\mathrm{star}\rangle=3.4\times10^{10}$\,M$_\odot$,
respectively.
The mean SFR for the excluded objects is also smaller 
than that of the objects shown in the plot. 
Therefore by excluding objects with bad SED fitting quality, 
it is possible that we miss objects that are located
lower left of the SFR-stellar mass diagram,  
and constitute low mass range of star-forming main sequence.

In each panel, representing different redshift bins, 
objects that lie above the thin line
(the limit where the SFRs of galaxies is 3 times 
that of the main sequence) are generally considered as 
starbursts \citep[e.g.,][]{Barrufet2020}. 
The fraction of objects that are classified as starbursts, 
$f_\mathrm{burst}$,
is defined as $N_\mathrm{burst}/N_\mathrm{total}$.
The $f_\mathrm{burst}$ is an indirect measure
of how much star formation quantity is 
induced by powerful events with short timescales
(such as gas-rich major mergers) 
or powered by relatively steady fueling mechanism
(such as continuous gas accretion). 
In the bottom right panel of Fig.~\ref{fig:sfrsmass},
we show how $f_{\rm burst}$ for submillimetre sources
changes as a function of redshift.
While the filled circles are from 193 objects with 
well constrained photometric redshifts ($\Delta z/(1+z)\le0.2$), 
the fainter, larger circles (without error bars) are from 
all 272 objects with good SED fitting quality.
The two values are comparable within the Poisson errors.
At $z\simeq2$, around 40 per cent 
of submillimetre sources are classified as starbursts. 
There is no clear trend of decreasing or increasing 
burst fraction with redshift,
if we use 193 objects with small uncertainties 
in photometric redshift. 
However, with the inclusion of $z\ge3$ objects 
that are mostly faint in wavelengths shortward of 3.6\,$\mu$m, 
the fraction of galaxies with bursting star formation 
is lower at $z\ge3$ compared to the case at $z<3$.
The burst fraction numbers do not change much even if
we include objects with less acceptable SED fitting quality 
(which are missed in the plot): 
$50\pm11$ per cent and $35\pm16$ per cent at $z<3$ and $z\ge3$.
\citet{Barrufet2020} have suggested that
the fraction of starbursts is 43 and 40 per cent
in bins of $1.5\le z<2.5$ and $2.5\le z<3.5$, respectively,
for dusty galaxies selected from \textit{Herschel}/SPIRE.
Using the same redshift bins, 
we obtain 52 and 44 per cent 
(or 44 and 41 per cent when no photometric redshift uncertainty cut is applied) 
for the burst fraction. 
The $f_{\rm burst}$ values are consistent with each other, 
suggesting that both SPIRE-selected and SCUBA-2-selected 
samples represent dusty star-forming galaxies at $z>2$,
while about 50 per cent of them are closely located on the main sequence
and another 50 per cent show largely enhanced star formation.

\subsection{AGN contribution and star formation}
\label{sec:AGN}

\begin{figure}
  \includegraphics[width=1.0\columnwidth]{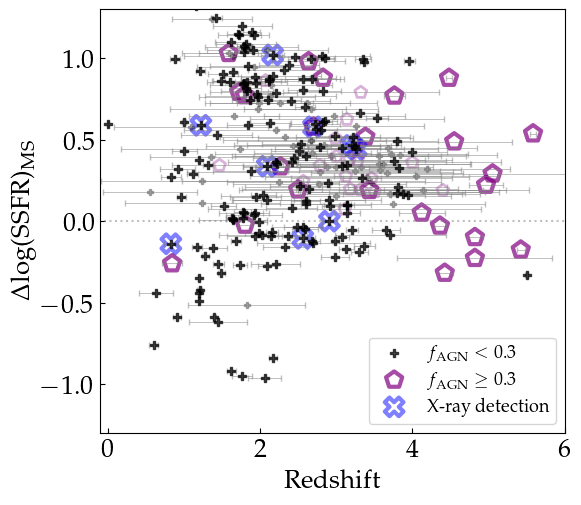}
  \caption{
	  Offset of the specific SFR from the
          star-forming galaxy main sequence, i.e.,
          $\Delta\mathrm{log(SSFR)}_\mathrm{MS}$,
	  of submillimetre sources 
          as a function of redshift.
	  Crosses and pentagons represent objects with 
	  $f_\mathrm{AGN}<0.3$ and $f_\mathrm{AGN}\ge0.3$, respectively,
	  i.e., more than 30 per cent of the infrared luminosity 
	  for the latter group 
	  is expected to originate from AGN.
	  Objects with $\Delta z/(1+z) > 0.2$ 
	  are marked with smaller (and fainter) symbols.
	  Submillimetre sources 
	  that are matched to the X-ray source catalogue 
	  are marked as crosses. Note that the SED fitting results for 
	  these X-ray identified sources 
	  indicate their $f_\mathrm{AGN}$ to be less than 0.3. 
        }
  \label{fig:sfagn}
\end{figure}

The AGN contribution fraction $f_\mathrm{AGN}$,
i.e., the ratio of the AGN luminosity
to the sum of the AGN and dust luminosities,
$L_\mathrm{AGN}/(L_\mathrm{AGN}+L_\mathrm{dust})$,
estimated from the \textsc{cigale} SED fitting,
can be used to indicate 
the role of AGN in a specific object. 
Since the uncertainties in $f_\mathrm{AGN}$ 
values are large compared to other quantities 
(such as SFR and stellar mass; Fig.~\ref{fig:mocktest}), 
we use $f_\mathrm{AGN}$ 
in the classification of objects, 
not in the comparison of AGN activity levels 
among objects. 
Different studies introduce different criteria
to discriminate AGN candidates 
from star-formation dominated galaxies
through $f_\mathrm{AGN}$ values 
\citep[e.g.,][]{Serjeant2010, Shen2020, Wang2020}.
For example, \citet{Serjeant2010} have classified 
850-$\mu$m selected objects with $f_\mathrm{AGN}>0.3$ as AGN, 
considering the uncertainty of estimating the $f_\mathrm{AGN}$
from the broad-band photometry SED fitting. 
\citet{Shen2020} have divided radio selected objects into 
star-forming galaxies ($f_\mathrm{AGN}<0.1$) 
and AGN ($f_\mathrm{AGN}>0.1$), 
and \citet{Wang2020} have suggested $f_\mathrm{AGN}>0.3$ 
to define MIR-selected AGN candidates.
Here, we adopt a cut of $f_\mathrm{AGN}\ge0.3$ 
for object classification. 
Objects with $f_\mathrm{AGN}\ge0.3$ are defined as 
`high-$f_\mathrm{AGN}$' sample, 
and objects with $f_\mathrm{AGN}<0.3$ are defined as 
`low-$f_\mathrm{AGN}$' sample.

Figure~\ref{fig:sfagn} shows 
272 objects with good SED fitting quality, 
divided into two subgroups: high-$f_\mathrm{AGN}$ objects
and low-$f_\mathrm{AGN}$ objects.
The fraction of high-$f_\mathrm{AGN}$ objects 
is lower than 10 per cent at $z<4$,
yet at redshift $z\ge4$,
over 70 per cent of the objects show $f_\mathrm{AGN}$ 
larger than 0.3. 
We can conclude that the AGN 
must have played an important role 
in the growth of $z>4$ dusty star-forming galaxies. 
Interestingly, most objects with X-ray detection 
\citep{Krumpe2015} show $f_\mathrm{AGN} < 0.3$.
This suggests
two possibilities: first, that X-ray surveys and MIR surveys
may probe 
different populations of AGN, 
since the MIR surveys have the power to reveal dusty AGN; 
second, the luminosity of X-ray selected AGN 
may account for only a small fraction of the infrared luminosity 
\citep{Pope2008}.
Further study on the $f_\mathrm{AGN}\ge0.3$ objects
will help decide which explanation is correct.

The enhancement of the SFR is parameterized
as the distance of a galaxy 
from the star-forming galaxy main sequence (MS)
in the SFR-stellar mass plane
(i.e., $\Delta\mathrm{log\,(SSFR)}_\mathrm{MS} = \mathrm{log}\,[\mathrm{SSFR_{galaxy}/SSFR_{MS}} (M_\mathrm{star},~z)]$).
By using this parameter, 
the effects of different stellar mass and redshift evolution are removed, 
and galaxies of all types of star formation
from the main sequence ($\Delta\mathrm{log\,(SSFR)}_\mathrm{MS}$ close to 0)
to vigorous starbursts
($\Delta\mathrm{log\,(SSFR)}_\mathrm{MS}$ significantly above 0)
can be distinguished. 
Previously in Fig.~\ref{fig:sfrsmass},
we show that objects with 
$\Delta\mathrm{log\,(SSFR)}_\mathrm{MS}>\mathrm{log}_{10}3$,
classified as starbursts, 
comprise about 40 per cent of the submillimetre sources, 
with a slight decrease at redshifts higher than $z\simeq3$. 
In Fig.~\ref{fig:sfagn}, 
average $\Delta\mathrm{log\,(SSFR)}_\mathrm{MS}$ values 
of submillimetre sources are about 0.34, 
suggesting that 850-$\mu$m SMGs are 
typically galaxies with twice as much SFR than 
normal star-forming galaxies at similar stellar mass. 
The mean $\Delta\mathrm{log\,(SSFR)}_\mathrm{MS}$ values 
are 0.35 and 0.34 for objects with 
$f_\mathrm{AGN}\ge0.3$ and $f_\mathrm{AGN}<0.3$, respectively,
based on 193 objects with small photometric redshift uncertainties. 
If objects with large photometric redshift uncertainties are included, 
the mean $\Delta\mathrm{log\,(SSFR)}_\mathrm{MS}$ values 
are 0.36 and 0.44 for high- and low-$f_\mathrm{AGN}$ objects.
The two-sample Kolmogorov-Smirnov test 
on the $\Delta\mathrm{log\,(SSFR)}_\mathrm{MS}$ values 
for high- and low-$f_\mathrm{AGN}$ subsamples shows 
$p$-values as large as 0.26
(when objects with large photometric redshift uncertainties are included)
and 0.98 (when only objects with 
small photometric redshift uncertainties are considered),
suggesting that the star formation properties 
for objects in two subsamples are not different.
This implies that 
an increase of AGN luminosity to total luminosity 
does not result in either a higher specific SFR
(increase in $\Delta\mathrm{log\,(SSFR)}_\mathrm{MS}$)
or a lower specific SFR
(decrease in $\Delta\mathrm{log\,(SSFR)}_\mathrm{MS}$).
This result is consistent with
previous studies on X-ray selected AGN,
which reported little evidence of
SFR being either enhanced or suppressed
by the increasing X-ray luminosity
\citep[e.g.,][]{Mullaney2012, Stanley2015, Ramasawmy2019}.

\subsection{Radio-to-FIR ratio}
\label{sec:xrayradio}

\begin{figure}
  \includegraphics[width=\columnwidth]{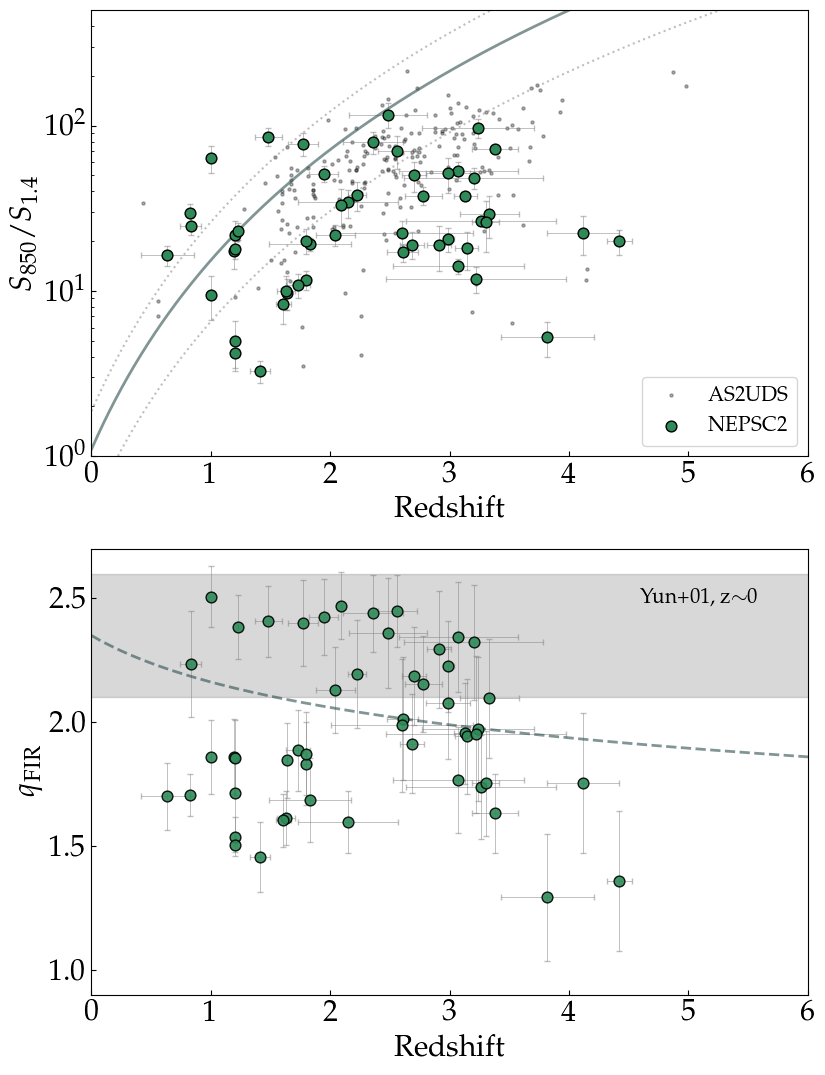}
  \caption{
	  {\it Top}: Submillimetre-to-radio flux density ratio
	  ($S_{850}/S_{1.4}$) versus
	  photometric redshift of radio-identified 
	  submillimetre sources. 
	  Small dots represent 
	  submillimetre sources identified in 
	  a higher spatial resolution submillimetre image 
	  \citep[][scaled to $S_{850}$ from $S_{870}$]{Dudzeviciute2020}.
	  The solid line illustrates the redshift dependence of 
	  $S_{850}/S_{1.4}$ presented in 
	  \citet{Barger2000}. 
	  The two dotted lines show the same trend with 
	  different dust temperatures, 80\,K and 20\,K.
	  {\it Bottom}: FIR-to radio correlation coefficient 
	  $q_\mathrm{FIR}$ versus photometric redshifts,
	  with both parameters derived from \textsc{cigale} SED fitting. 
	  The grey shaded region indicates the range 
	  $q_\mathrm{FIR}=2.34\pm0.25$, 
	  which covers the values seen for local FIR-selected galaxies
	  \citep{Yun2001}. The dashed line is the suggested 
	  redshift evolution of the FIR-to-radio correlation 
	  for main-sequence galaxies \citep{Magnelli2015}, 
	  based on \textit{Herschel} photometry and a stacking analysis
	  out to $z=2.5$.  
        }
  \label{fig:radio}
\end{figure}

SED fitting of radio-identified submillimetre counterparts 
is carried out by including the VLA photometry.
It has been suggested that 
the observed radio spectral index 
between 1.4\,GHz and 350\,GHz (i.e., 850\,$\mu$m)
can be used in the estimation of photometric redshifts, 
based on the tight radio-to-FIR correlation
\citep[e.g.,][]{Barger2000, Dunlop2001, YunCarilli2002, Aretxaga2003}.
Figure~\ref{fig:radio} (top) shows 
the 850-$\mu$m to 1.4\,GHz flux ratio
$S_{850}/S_{1.4}$ 
of the radio-identified submillimetre sources 
versus the estimated photometric redshift, found by SED fitting. 
We exclude submillimetre sources with 
multiple counterparts in this plot, 
since the 850-$\mu$m flux density might include contributions from
more than one object in those cases.
Cases where the SED fitting quality is not acceptable 
are also excluded.

In the $S_{850}/S_{1.4}$ versus redshift plot,
850-$\mu$m selected submillimetre sources 
from the NEPSC2 occupy a similar region 
as the submillimetre sources selected in the higher spatial resolution
870-$\mu$m image \citep[AS2UDS;][]{Dudzeviciute2020}.
This suggests that the
850-$\mu$m flux density measured in NEPSC2 
is not contaminated by a radio-quiet component. 
The solid line plotted for comparison is the expected trend for 
$S_{850}/S_{1.4}$ as a function of redshift, 
based on the local infrared-luminous galaxy Arp 220 
\citep{Barger2000}. 
In their work, the SED of Arp 220 over 30--3000\,$\mu$m 
is approximated as a modified blackbody 
with emissivity $\beta=1$ 
and a dust temperature of 47\,K. 
By assuming the power-law slope of the synchrotron emission 
to be 0.8, they suggest that $S_{850}/S_{1.4}$  
is proportional to $(1+z)^{3.8}$ 
and is higher at larger dust temperatures.
In the figure, 
two dotted lines represent the same trend 
with dust temperatures of 80\,K and 20\,K. 
The majority of the radio-identified 
submillimetre sources lie below the \citet{Barger2000} line,
implying lower dust temperature 
and/or different $\beta$. 
There are sources at redshifts around $z\simeq4$
with low $S_{850}/S_{1.4}$ ($<30$), i.e., 
excess radio emission. 
Two of these objects have no detection in optical images,
while the remaining one is very faint, 
implying that these may be candidates for heavily attenuated 
radio-loud AGN at high redshifts. 

Rather than using the observed flux density ratio, 
the FIR-to-radio correlation coefficient 
\citep[][$q_\mathrm{FIR}$]{Helou1985},
derived using the IR luminosity from full SED fitting,
would be a better parameter that can be compared
among different galaxy populations. 
The $q_\mathrm{FIR}$ values for 
radio-identified submillimetre sources 
are derived
using \textsc{cigale} SED fitting, 
with the determination coefficient $r^2=0.84$ 
based on the mock tests (Fig.~\ref{fig:mocktest}). 
In the bottom panel of Fig.~\ref{fig:radio}, 
we show the $q_\mathrm{FIR}$ values 
along the redshift axis, 
for the same sources presented in the top panel.
Local galaxies selected in the FIR 
are known to show
very consistent $q_\mathrm{FIR}\simeq2.3$ values \citep{Yun2001}. 
The radio-identified submillimetre sources in our sample
show $\langle q_\mathrm{FIR}\rangle=2.0\pm0.3$, 
lower than the value for local IR luminous galaxies. 
The change in the $\langle q_\mathrm{FIR}\rangle$ value 
in the redshift range of $z=2$--3 
is roughly consistent with the 
evolution of FIR-to-radio correlation 
out to $z=2.5$, 
inferred from a stacking analysis of 
star-forming main-sequence galaxies
\citep{Magnelli2015}.
While the radio-identified submillimetre sources 
show different levels of star formation enhancement, 
the consistency between our $\langle q_\mathrm{FIR} \rangle$ value
and that of \citet{Magnelli2015} confirms that the
FIR-to-radio correlation is still 
universal, regardless of the star-formation mode
of a galaxy.

Objects with small $q_\mathrm{FIR}$ are classified as
`radio-excess' objects, i.e., possible radio-loud AGN. 
We notice that there exist objects clustered at 
$z\simeq1.5$ and $z\simeq4$ with low $q_\mathrm{FIR}$, 
which are also distinguished from others
because of the low $S_{850}/S_{1.4}$ values. 
The fraction of such radio-excess objects 
among the radio-identified submillimetre sources 
is about 15\,per cent.
These radio-excess objects need further study,
particularly through obtaining spectroscopic redshifts.

\section{Conclusions}
\label{sec:summary}

We have presented the multi-wavelength counterparts of 
850-$\mu$m sources in the NEP survey fields 
along with their physical properties 
derived from SED fitting. 
By using the 20-cm radio flux density measured in a higher resolution image,
and optical-to-NIR colours, 
we identify 514 counterparts to 449 submillimetre sources. 
The identification rate is 69 per cent (449/647). 
The optical-to-NIR colour identification recovers 
84 per cent of the radio counterparts,
showing that the robustness of using colours 
in identification of submillimetre source counterparts. 
The number fraction of 850-$\mu$m sources
with more than one identified counterpart 
is about 10 per cent.
Our main conclusions
on the nature of 850-$\mu$m selected sources (i.e., SMGs)
found through the SED fitting are as follows.

(i) The mean photometric redshift of the SMGs 
is $\langle z \rangle=2.5$, similar to values from
previous studies of 850-$\mu$m-selected galaxies. 
The derived photometric redshifts 
are consistent with the spectroscopic redshifts, 
despite the number of objects with available
spectroscopic redshifts being small.
The total infrared luminosities 
are approximately comparable to typical values for local ULIRGs, 
ranging over $\mathrm{log}_{10} (L_\mathrm{IR}/\mathrm{L}_\odot)=11.5$--13.5. 
The NEPSC2 SMG sample includes a number of optically faint SMGs,
with a mean redshift of $\langle z \rangle=3.2$.

(ii) In the SFR-stellar mass diagrams
for different redshift bins, 
at least 40 per cent of the SMGs are classified as bursty systems 
whose SFRs are more than 3 times larger 
than the star-forming main sequence galaxies at those redshifts.
If the samples are limited to those with 
smaller photometric redshift uncertainty, 
no clear trend is found for starburst fraction
varies along the redshift.
However, if we consider the existence of 
optically undetected or very faint dusty galaxies 
at $z\ge3$ that lie on the main sequence, 
the burst fraction appears to decrease 
as redshift increases.
It is not straightforward to suggest a physical mechanism 
to explain the SFRs of these galaxies 
from this result;
however, there must be some process 
that can explain the `less bursty' mode of star formation 
in these high-redshift dusty galaxies.

(iii) At $z\ge4$, most objects that are thought to be 
responsible for submillimetre emission 
are classified as high-$f_\mathrm{AGN}$ objects,
i.e., objects with AGN contribution to the total infrared luminosity 
larger than 0.3. 
The SFR enhancement, defined as an offset from 
the star-forming main sequence, 
for high-$f_\mathrm{AGN}$ and low-$f_\mathrm{AGN}$ objects
is not statistically different. 
This shows that AGN play an important role
at early times for dusty star-forming galaxy growth, 
yet there is no clear evidence that 
the presence of AGN either enhances or suppresses the SFR
in the galaxy.

(iv) The scatter in the FIR-to-radio correlation coefficients
$q_\mathrm{FIR}$
for submillimetre counterparts is relatively large.
While we see $q_\mathrm{FIR}$ values at $2<z<3$
that are similar to those
expected from redshift evolution, 
some objects show even larger radio excess. 
This diagnostic could be used in future to
search for radio-loud AGN at high redshift,
and we already have possible candidates that are worth following up.

The investigation of the nature of 850-$\mu$m selected galaxies
with counterpart identification in this paper 
could be expanded further 
using the upcoming data sets; 
for example, deep X-ray observations from 
the \textit{eROSITA} survey 
\citep[][]{Merloni2020} 
and millimetre observations using the 
LMT/TolTEC 
\citep[][3-bands survey with 1.1, 1.4, and 2.0\,mm]{Bryan2018} 
will add more constraints on the AGN activity 
and dust characteristics of these sources. 
Connections between SMGs and galaxy large-scale structure
will also be investigated 
with the help of planned or ongoing spectroscopic surveys 
on the NEP region, using MMT/Hectospec and Subaru/PFS.

\section*{Data availability}

Supplementary data are available at MNRAS online. 

\noindent Table A1. 
Counterpart identification results for entire 850-$\mu$m sources 
from the NEPSC2$+$S2CLS survey. 
As the counterpart identification has not been made for 
three sources that are located close to a saturated star 
or a Galactic planetary nebula, 
the number of 850-$\mu$m sources presented in this table is 644. 

\noindent Table A2. Catalogue containing the multi-wavelength photometry 
of the 514 identified counterparts 
(either with radio or with colour identification that are reliable) 
and the \textsc{cigale} outputs for each object. 
The outputs include photometric redshift, 
total IR luminosity, SFR, stellar mass, $A_V$, $f_\mathrm{AGN}$, 
and $q_\mathrm{FIR}$, as well as their uncertainties.

\section*{Acknowledgements}

The James Clerk Maxwell Telescope is operated by
the East Asian Observatory on behalf of:
The National Astronomical Observatory of Japan;
Academia Sinica Institute of Astronomy and Astrophysics;
the Korean Astronomy and Space Science Institute;
and the Center for Astronomical Mega-Science (as well as the National
Key R\&D Program of China with No. 2017YFA0402700).
Additional funding support is provided by the
Science and Technology Facilities Council of the
United Kingdom and participating universities in the
United Kingdom and Canada.
Additional funds for the construction of SCUBA-2 were
provided by the Canada Foundation for Innovation.
This paper uses the JCMT data 
taken as part of Program ID M17BL007, M20AL005, 
and archival data from Program ID MJLSC02.
HS, DL and YK acknowledge support from the
National Research Foundation of Korea (NRF) grant,
No. 2018R1C1B6008498 and No. 2021R1A2C4002725,
funded by the Korea government (MSIT).
DS acknowledges support from the Natural Sciences and Engineering Council of Canada.
SS was supported in part by ESCAPE - The European Science Cluster of Astronomy \& Particle Physics ESFRI Research Infrastructures, which in turn received funding from the European Union's Horizon 2020 research and innovation programme under Grant Agreement no. 824064. SS also thanks the Science and Technology Facilities Council for support under grant ST/P000584/1.
YA acknowledges financial support by NSFC grant 11933011 and the science research grant from the China Manned Space Project with NO. CMS-CSST-2021-B06.
TG acknowledges the support by the Ministry of Science and Technology of Taiwan through grant105-2112-M-007-003-MY3.
HSH acknowledges support by the National Research Foundation of Korea (NRF) grant funded by the Korea government (MSIT) (No. 2021R1A2C1094577).
MI acknowledges the support from the National Research Foundation
of Korea (NRF) grant, No. 2020R1A2C3011091, funded by
the Korea government (MSIT). 
MPK acknowledges support from the First TEAM grant 
of the Foundation for Polish Science No. POIR.04.04.00-00-5D21/18-00.
MJM acknowledges the support of
the National Science Centre, Poland through the SONATA BIS grant 2018/30/E/ST9/00208.
GJW gratefully acknowledges an Emeritus Fellowship from 
The Leverhulme Trust.











\setlength{\tabcolsep}{3pt}
\begin{landscape}

\begin{table}
\scriptsize
\caption{Selected entries from the counterpart identification results for 850-$\mu$m sources around the NEP area}
   \label{tab:coord}
       \begin{tabular}{lclllcccccccccccccc}
         \hline
         \hline
	       SMM ID & NEPSC2 ID & SC2CLS ID & Radio ID & Opt/IR ID & RA (850\,$\mu\mathrm{m}$) & Dec (850\,$\mu\mathrm{m}$) & RA (radio) & Dec (radio) & RA (IRAC) & Dec (IRAC) & RA (optical) & Dec (optical) & $S_{850}$ & $\sigma_{850}$ & radio & colour & reliable & multiple \\
         \hline
	       2   & NEPSC2\_J175052+660458 & --- & --- &  2-1 & 267.7205 &  66.0828 &       --- &       --- & 267.71922 &  66.08245 & 267.71855 &  66.08249 &  32.039 & 3.139 & N & Y & Y & N \\
	       10  & NEPSC2\_J180457+671542 & --- & --- & 10-1 & 271.2404 &  67.2617 &       --- &       --- & 271.24429 &  67.26094 & 271.24402 &  67.26064 &  10.577 & 4.066 & N & Y & N & N \\
	       20  & NEPSC2\_J180157+671747 & --- & --- & 20-1 & 270.4882 &  67.2966 &       --- &       --- & 270.48607 &  67.29675 & 270.48616 &  67.29679 &   9.424 & 3.421 & N & Y & Y & Y \\
	       20  & NEPSC2\_J180157+671747 & --- & --- & 20-2 & 270.4882 &  67.2966 &       --- &       --- & 270.49095 &  67.29782 & 280.49121 &  67.29783 &   9.424 & 3.421 & N & Y & Y & Y \\
	       23  & NEPSC2\_J175619+664606 & NEP.0003  & 23a   & 23-21 & 269.0807 &  66.7686 & 269.07967 &  66.76822 & 269.07975 &  66.76818 & --- &  --- & 11.288 & 1.805 & Y & Y & Y & Y \\
	       23  & NEPSC2\_J175619+664606 & NEP.0003  & 23b   & 23-22 & 269.0807 &  66.7686 & 269.08212 &  66.76803 & 269.08219 &  66.76800 & 269.08220 & 66.76815 & 11.288 & 1.805 & Y & Y & Y & Y \\
	       29  & NEPSC2\_J175515+664355 & NEP.0010 & 29   & 29-1 & 269.8131 &  66.7321 & 268.81263 &  66.73238 & 268.81288 &  66.73238 & 268.81291 &  66.73248 & 10.963 & 1.782 & Y & Y & Y & N \\
	       63  & NEPSC2\_J175351+670606 & --- & ---  & 63-1 & 268.4652 &  67.1018 &       --- &       --- & 268.46617 &  67.10250 &       --- &       --- & 7.904 & 2.850 & N & Y & Y & N \\
	       100 & NEPSC2\_J180006+663654 & NEP.0017 & ---  &100-1 & 270.0286 &  66.6151 &       --- &       --- & 270.03033 &  66.61509 & 270.03014 &  66.61510 & 8.339 & 1.857 & N & Y & Y & N \\
	       163 & NEPSC2\_J175236+663813 & NEP.0015 & 163a &163-1 & 268.1514 &  66.6371 & 268.15002 &  66.63690 & 268.14993 &  66.63681 & 268.14987 &  66.63675 & 7.272 & 1.816 & Y & Y & Y & Y \\
	       163 & NEPSC2\_J175236+663813 & NEP.0015 & 163b &163-3 & 268.1514 &  66.6371 & 268.15049 &  66.63855 & 268.15078 &  66.63841 & 268.15084 &  66.63848 & 7.272 & 1.816 & Y & N & N & Y \\
	       282 & NEPSC2\_J175239+662327 & --- & 282  & ---  & 268.1650 &  66.3910 & 268.16328 &  66.39084 &       --- &       --- &       --- &       --- &  5.324 & 1.777 & Y & N & N & N \\
	       470 & NEPSC2\_J175438+664303 & NEP.0125 & ---  &470-1 & 268.6591 &  66.7177 &       --- &       --- & 268.65764 &  66.71719 & 268.65960 &  66.71797 & 3.696 & 1.825 & N & Y & N & Y \\
	       470 & NEPSC2\_J175438+664303 & NEP.0125 & 470a &470-31& 268.6591 &  66.7177 & 268.66182 &  66.71960 & 268.66198 &  66.71960 & 268.66234 &  66.71958 & 3.696 & 1.825 & Y & N & N & Y \\
	       470 & NEPSC2\_J175438+664303 & NEP.0125 & 470b &470-32& 268.6591 &  66.7177 & 268.66233 &  66.71523 & 268.66246 &  66.71523 & 268.66256 &  66.71526 & 3.696 & 1.825 & Y & N & N & Y \\
	       20050 & --- & NEP.0050 & NEP.0050a & 20050-3 & 269.6466 &  66.6166 & 269.64728 &  66.61698 & 269.64711 &  66.61717 & 269.64714 &  66.61711 &  5.286 & 1.399 & Y & N & N & Y \\
	       20050 & --- & NEP.0050 & NEP.0050b & 20050-1 & 269.6466 &  66.6166 & 269.64491 &  66.61607 & 269.64550 &  66.61618 & 269.64545 &  66.61611 &  5.286 & 1.399 & Y & Y & Y & Y \\
	       20092 & --- & NEP.0092 & NEP.0092  & 20092-1 & 268.6824 &  66.4739 & 268.67864 &  66.47413 & 268.67893 &  66.47407 & 268.67873 &  66.47396 &  4.526 & 1.318 & Y & Y & N & N \\
	       20327 & --- & NEP.0327 & ---  & 20327-1 & 269.0750 &  66.4985 & 269.07075 &  66.49810 & 269.07102 &  66.49809 & 269.07080 &  66.49812 &  2.973 & 1.180 & N & Y & Y & Y \\
	       20327 & --- & NEP.0327 & NEP.0327  & 20327-3 & 269.0750 &  66.4985 & 269.07075 &  66.49810 & 269.07102 &  66.49809 & 269.07080 &  66.49812 &  2.973 & 1.180 & Y & N & N & Y \\
         \hline
       \end{tabular}
	\vspace{1ex}

	{\raggedright \textit{Note.}
	Column (1): 850-$\mu$m source ID in the combined catalogue. 
	Column (2): 850-$\mu$m source ID from the NEPSC2 catalogue \citep{Shim2020}.
	Column (3): 850-$\mu$m source ID (`Nickname') from the S2CLS catalogue \citep{Geach2017}. 
	Column (4): radio counterpart ID presented in Section \ref{sec:radio-id},
	with suffices `a' and `b' in order of positional offsets between the 
	850-$\mu$m position and 20-cm position. 
	Column (5): counterpart ID presented in Section \ref{sec:colour-id},
	where `-1' and `-2' stands for primary and secondary counterpart 
	from the optical-NIR colour identification. IDs with `-3' indicates that 
	the object is not originally identified from the optical-NIR colour 
	identification, because of the low relative possibility $p_i$.
	If there is more than one number in the suffices, e.g. `-21' and `-22', 
	it means that more than one secondary counterpart is identified
	because of the definition of the secondary identification. 
	Columns (6) and (7): 850-$\mu$m coordinates.
	Columns (8) and (9): coordinates of the counterparts in the VLA images. 
	Columns (10) and (11): coordinates of the counterparts in the IRAC 4.5\,$\mu$m. 
	Columns (12) and (13): coordinates of the counterparts in the HSC $g$-band
	images (when available) or in the $z$- or $y$-band images (when the object is 
	undetected in the $g$ band). 
	Columns (14) and (15): flux density and uncertainty in 850\,$\mu$m, in units of mJy. 
	Column (16)--(19): boolean column 
	indicating whether the counterpart is identified in radio wavelength (20\,cm),
	whether the counterpart is identified using optical-NIR colour, 
	whether the identification based on colour is reliable enough 
	(with false rate less than 10 per cent), 
	and whether the submillimetre source has more than one counterpart. \par}
\end{table}

\end{landscape}


\bsp	
\label{lastpage}
\end{document}